\documentclass[10pt, journal]{IEEEtran}
\usepackage{pgfplots}
\usepackage{tikz}
\usepackage{tikzpeople}
\usepackage{circuitikz}
\usetikzlibrary{tikzmark, shapes, fit,backgrounds}
\usetikzlibrary{matrix, decorations.pathreplacing, angles, quotes}
\usetikzlibrary{patterns, calc, positioning}
\usetikzlibrary{arrows, arrows.meta}
\usetikzlibrary{shapes.multipart}
\usetikzlibrary{fit}
\usetikzlibrary{shapes.geometric, positioning}
\usetikzlibrary{decorations.pathmorphing}
\tikzstyle{block} = [rectangle, draw, 
    minimum width=4em, text centered, rounded corners, minimum height=1em]
\tikzstyle{line} = [draw, -latex]
\tikzset{meter/.append style={draw, inner sep=10, rectangle, font=\vphantom{A}, minimum width=30, scale=.7, path picture={\draw[black] ([shift={(.1,.3)}]path picture bounding box.south west) to[bend left=50] ([shift={(-.1,.3)}]path picture bounding box.south east);\draw[black,-{Latex[scale=.5]}] ([shift={(0,.1)}]path picture bounding box.south) -- ([shift={(.3,-.1)}]path picture bounding box.north);}}}
\tikzset{snake it/.style={decorate, decoration=snake}}\def\bbsmatrix#1{\begin{bsmallmatrix}#1\end{bsmallmatrix}}

\usepackage{pgfplots}
\usetikzlibrary{shapes.multipart}
\usetikzlibrary{fit}
\usetikzlibrary{shapes.geometric,positioning}
\usetikzlibrary{calc}
\usepackage{colortbl}
\usetikzlibrary{patterns}
\usetikzlibrary{matrix}
\usetikzlibrary{tikzmark}
\usepackage{blkarray}
\usepackage{booktabs}
\usepackage{soul}
\usepackage{arydshln}
\usetikzlibrary{patterns.meta}
\usepackage{caption}
\usepackage{subcaption}
\usepackage{url}
\usepackage{hyperref}
\usepackage{stfloats}
\usepackage{braket}
\usepackage{booktabs}
\usepackage{siunitx}
\usepackage{bm}
\usepackage[utf8]{inputenc}
\usepackage{mathtools}
\usepackage{graphicx}
\usepackage{algorithm}
\usepackage{cite}
\usepackage{amsmath,amssymb,amsfonts, amsthm, mathrsfs}
\usepackage{algorithmic}
\usepackage{textcomp}
\usepackage{xcolor}
\usepackage{enumitem}

\tikzset{
    pics/mybox/.style n args={3}{
        code={
    \draw[help lines](#1,0,0)--(#1,#2,0);
    \draw[help lines](0,#2,0)--(#1,#2,0);
    \draw[help lines](0,#2,0)--(0,#2,#3);
    \draw[help lines](0,0,#3)--(0,#2,#3);
    \draw[help lines](#1,0,0)--(#1,0,#3);
    \draw[help lines](0,0,#3)--(#1,0,#3);
   \draw[help lines](#1,#2,0)--(#1,#2,#3);
   \draw[help lines](0,#2,#3)--(#1,#2,#3);
      \draw[help lines](#1,0,#3)--(#1,#2,#3);
}}}

\newtheorem{theorem}{Theorem}
\newtheorem{example}{Example}
\newtheorem{remark}{Remark}

\newtheorem{definition}{Definition}

\DeclareMathOperator*{\argmax}{argmax}

\DeclareMathOperator*{\Tr}{Tr}
\DeclareMathOperator*{\conv}{conv}
\DeclareMathOperator*{\diag}{diag}

 \def\ppsmatrix#1{\begin{psmallmatrix}#1\end{psmallmatrix}}

\allowdisplaybreaks

\title{
On the Utility of Quantum Entanglement  for Joint   Communication and Instantaneous Detection}
\author{
\IEEEauthorblockN{Yuhang Yao and Syed A. Jafar}\\
\IEEEauthorblockA{
Center for Pervasive Communications and Computing (CPCC)\\
University of California Irvine, Irvine, CA 92697\\
Email: \{yuhangy5, syed\}@uci.edu
}
}

\date{}      

\begin{document}
\maketitle

\IEEEpubid{\begin{minipage}{\textwidth}\ \\[40pt] \centering
  This work has been submitted to the IEEE for possible publication. \\ \centering Copyright may be transferred without notice, after which this version may no longer be accessible.
\end{minipage}}

\begin{abstract}
Entanglement is known to significantly improve the performance (\emph{separately}) of communication and detection schemes that utilize quantum resources. This work explores the simultaneous utility of quantum entanglement for (\emph{joint}) communication and  detection schemes, over channels that are convex combinations of identity, depolarization and erasure operators, both with perfect and imperfect entanglement assistance. The channel state is binary, rapidly time-varying and unknown to the transmitter. While the communication is delay-tolerant, allowing the use of arbitrarily long codewords to ensure reliable decoding, the channel state detection is required to be instantaneous. The detector is neither co-located with the transmitter, nor able to wait for the decoding in order to learn the transmitted waveform. The results of this work appear in the form of communication-rate vs instantaneous-detection-error tradeoffs, with and without quantum entanglement. Despite the challenges that place the two tasks at odds with each other, the results indicate that quantum entanglement can indeed be simultaneously and significantly beneficial for joint communication and instantaneous detection.
\end{abstract}

\allowdisplaybreaks

\section{Introduction} \label{sec:introduction}
Quantum technologies are increasingly attracting attention  beyond quantum computation, for their applications to communication, estimation and detection.  Due to the unique properties of quantum systems such as entanglement and superposition, protocols that take advantage of such properties can surpass their classical counterparts. For example, quantum entanglement can be used by quantum sensing (and metrology) schemes to improve the mean square error beyond the standard quantum limit \cite{Lloyd}; by quantum illumination (e.g., quantum radar) schemes to significantly improve the probability of detection error \cite{lloyd2008enhanced, shapiro2020quantum}; and by quantum communication schemes  for multiplicative gains in capacity via superdense coding \cite{bennett_shor_capacity}. 
 
The tasks of communication, estimation and detection are so closely related that they are oftentimes studied jointly. For example, integrated sensing and communication (ISAC) has been a highly active topic of research in wireless communication \cite{liu2022integrated,liu2022survey,zhang2011joint} and classical information theory \cite{Joudeh_Willems_2022, Wu_Joudeh, chang2023rate, ahmadipour2023information, Ahmadipour_Kobayashi_Wigger_Caire, xiong2023fundamental}. More recently, the topic is also beginning to be explored in the quantum setting \cite{wang2022joint, pereg2022estimation, liu2024quantumISAC}. Taking a step in this direction, in this work we explore the utility of quantum entanglement for \emph{joint  communication and instantaneous detection} (JCID). Both of these tasks seek information via a communication channel: the former (communication) focuses on coding to counter the channel imperfections in order to reliably recover a message generated by a source, whereas the latter (instantaneous detection) tries to learn an unknown parameter passively (uncoded) generated by nature, which may be considered as the instantaneous binary state (e.g., representing the presence/absence of a dynamic target) of a time-varying channel. In this work we are interested in settings where the time scales for the two tasks are quite different. Specifically, we consider the utility of quantum entanglement for joint optimization of delay-tolerant communication protocols that allow coding over arbitrarily long coding blocks, in conjunction with detection protocols that are highly delay-sensitive, namely \emph{instantaneous detection}. Instantaneous detection is particularly desirable in dynamic environments such as gaming, defense, and automated driving.

We focus on quantum channels that are convex combinations of \underline{i}dentity, \underline{d}epolarizing, and \underline{e}rasure operators --- IDE channels in short. These channels are among the most widely prevalent models in the literature due to their analytical tractability which often allows closed form expressions for key performance metrics. IDE channels are especially interesting for JCID because they include settings where some of the largest gains from quantum entanglement have been reported (separately) for both detection \cite{lloyd2008enhanced} and communication  \cite{bennett_shor_capacity}.  The detection problem considered by  Lloyd \cite{lloyd2008enhanced} under the quantum illumination framework distinguishes between two states of an IDE channel, by sending and receiving a quantum system (photon) with or without entanglement assistance. Under the single-photon model\footnote{The gain is a more modest $6$ dB over coherent states \cite{shapiro2020quantum}.} a factor of $d$ gain in SNR is established in  \cite{lloyd2008enhanced} due to quantum entanglement, by using $d$-mode entangled photon pairs. A comparable gain from entanglement is established for quantum communication by Bennett et al. in \cite{bennett_shor_capacity}, whereby together with the result of King \cite{king2003capacity}, it is shown that the (classical) communication capacity of the $d$-ary quantum depolarizing channel can be larger by a factor as large as $d+1$ relative to the capacity without entanglement assistance. Evidently, each of communication and detection can individually benefit significantly from quantum entanglement over IDE channels. Since entanglement is a precious resource, it is also important to understand whether (and to what extent) communication and detection can \emph{simultaneously} benefit from quantum entanglement. This is the motivation for our study of JCID over IDE channels in this work.

Given that studies of quantum systems for integrated communication, detection  and/or estimation are in their early stages, there is little existing work that is directly related to our problem formulation of JCID over IDE channels. Perhaps the closest to our setting is  \cite{wang2022joint} which studies the joint performance of detection and communication over a classical-quantum (c-q) channel with an unknown static channel parameter, obtaining a fundamental rate/detection-error exponent tradeoff  that depends on the empirical distribution of the codeword. However,  the differences from our setting are too substantial to allow meaningful comparisons. First, unlike our setting, the c-q channel model of \cite{wang2022joint} does not naturally lend itself to comparisons of performance with and without entanglement. This is primarily because if a c-q channel is given, there is no room to model the entanglement between the transmitter and the receiver since the quantum resource is only related to the receiver side, although it can be relaxed by, e.g., a comparison between two c-q channels that model the availability of entanglement in different ways. Second,  \cite{wang2022joint} assumes that the detector and transmitter are co-located, whereas in our setting  the detector is separate from the transmitter. The difference is significant because in the  setting of \cite{wang2022joint} the detector knows the transmitted signal waveform, whereas in our setting the detector must also overcome the uncertainty in the transmitted signal due to the communication task. As a result, the desired waveforms for communication and instantaneous detection are much more conflicted in our setting because  the detector would typically prefer a perfectly deterministic transmitted signal so that the uncertainty is only due to the unknown channel parameter, while the communication receiver would typically prefer a maximally random signal in order to carry as much source information as possible. Third, while the channel parameter in \cite{wang2022joint} is static, our setting considers a dynamic (e.g., an i.i.d. varying) channel state. Notably even in classical settings, static channel state models  \cite{Joudeh_Willems_2022,Wu_Joudeh,chang2023rate} pose a different set of challenges from dynamic channel models \cite{Ahmadipour_Kobayashi_Wigger_Caire,ahmadipour2023information}.\footnote{Dynamic channel states in non-i.i.d. settings such as \cite{nikbakht2024memory,lindstrom2025rate} where the channel states have more general correlations across channel uses. e.g., with Markov states \cite{lindstrom2025rate} tend to be even more challenging.} Furthermore, even when the channel model is dynamic, the requirement of \emph{instantaneous} detection  in JCID is significant, not only because fast detection may be important in practice, but because of its technical challenges. This is because  it prevents the detector from exploiting the long-term correlations  in the code structure. For example, if the detector could afford to wait for the decoding, then reliable decoding guarantees would essentially provide the detector the knowledge of the transmitted waveform, bringing our setting closer to the setting of \cite{wang2022joint}. Instantaneous detection does not allow this in our problem formulation.

The main results of this work appear in the form of communication-rate vs instantaneous-detection-error tradeoffs with and without quantum entanglement. The optimal tradeoff is characterized for the unentangled setting (Theorem \ref{thm:unentangled}), while an achievable tradeoff is established for entangled protocols with perfect (Theorem \ref{thm:superdense_coding}) or imperfect (Theorem \ref{thm:unreliable}) entanglement assistance. Taking advantage of the tractability of IDE channels, these tradeoffs are presented in explicitly computable forms that facilitate numerical evaluations of the utility of quantum entanglement. These evaluations (see Examples \ref{ex1}, \ref{ex2}, \ref{ex3}) support the conclusion that quantum entanglement can indeed be simultaneously beneficial for communication and instantaneous detection protocols towards significant improvements in their respective performance metrics (rate and weighted error probability).

\vspace{0.2cm}

\noindent {\it Notation:} $\mathbb{N}$ denotes the set of positive integers. $\mathbb{Z}$ denotes the set of integers. For an integer $d>1$, $\mathbb{Z}_d \triangleq \mathbb{Z}/d\mathbb{Z}$, i.e., integers modulo $d$. For $a \in \mathbb{N}$, $[a]$ denotes the set of integers $\{1,2,\cdots, a\}$. $[a:b]$ denotes the set of positive integers $\{a,a+1,\cdots, b\}$. For  $S\subseteq \mathbb{N}$, the notation $x_S\triangleq \{x_i: i\in S\}$.  The  convex hull of a set $S$ is denoted as $\conv S$. $\mathcal{H}_d$ represents the $d$-dimensional Hilbert space. $I_d$ denotes the $d\times d$ identity matrix. $\mathcal{D}(\mathcal{H})$ denotes the space of density operators acting on $\mathcal{H}$.  $\Pr(E)$ is the probability of an event $E$. 
$P_{\mathcal{X}}$ denotes the set of all probability distribution functions (vectors) on the set $\mathcal{X}$, i.e., $P_X\in P_{\mathcal{X}}$ defines a random variable $X$ such that $P_X(x) \triangleq \Pr(X=x)$ for $x\in \mathcal{X}$. With a little abuse of notation, we let ${\sf H}(X) = {\sf H}(P_X)$ denote the Shannon entropy with respect to a random variable $X$, or with respect to a distribution $P_X$. Similarly for a quantum system $A$ with its density operator $\rho_{A}$, ${\sf H}(A)_{\rho_{A}} = {\sf H}(\rho_A)$ denotes its von-Neumann entropy. We define $x\log_2 (x) \triangleq 0$ for $x=0$. Conditional entropy and mutual information are defined in the conventional way, e.g., as in \cite{Wilde_2017}.
For a joint classical quantum system $XA$ where $X$ is classical and $A$ is quantum, with joint density operator $\rho_{XA}$, $\rho_{A\mid X=x}$ denotes the density operator for $A$ conditioned on $X=x$. It is also written compactly as $\rho_{A\mid x}$ when the classical random variable in the condition is clear from the context.
$\Vert M \Vert_1 = \Tr(\sqrt{{M^\dagger M}})$ denotes the trace norm (Schatten 1-norm) of an operator $M$. By definition, the trace norm is equal to the sum of singular values of  $M$, and thus the sum of the absolute values of its eigenvalues if $M$ is a normal matrix. For an $m_1\times n_1$ matrix $A$, and an $m_2\times n_2$ matrix $B$, define the generalized sum operator $\boxplus$ so that $C= A\boxplus B$ is the $\max(m_1,m_2) \times \max(n_1,n_2)$ matrix obtained by the (regular) addition of $A'$ and $B'$, i.e., $C=A'+B'$, where $A'$ and $B'$ are $\max(m_1,m_2) \times \max(n_1,n_2)$ matrices obtained from $A$ and $B$ respectively by appending as many all-zero columns to the right and/or all-zero rows to the bottom as needed to satisfy the expanded matrix dimensions. For example, $\ppsmatrix{1\\ 2}~\boxplus~\ppsmatrix{3&4}~\boxplus~\ppsmatrix{5}=\ppsmatrix{1&0\\ 2&0}+\ppsmatrix{3&4\\0&0}+\ppsmatrix{5&0\\0&0}$.

\section{Joint Communication and Instantaneous Detection (JCID)} \label{sec:problem}
To study joint communication and instantaneous detection over quantum channels, with a focus on the entanglement between the transmitter and the receiver, let $\mathcal{N}^{(1)}, \mathcal{N}^{(2)}$ be two quantum channel states  such that $\mathcal{N}^{(s)} \colon \mathcal{D}(\mathcal{H}_{d}) \mapsto \mathcal{D}(\mathcal{H}_{d'})$ is a CPTP map for $s\in \{1,2\}$. Let $\mathcal{N} \triangleq \theta_1 \mathcal{N}^{(1)} + \theta_2 \mathcal{N}^{(2)}$ be the overall channel.
At time slot $t\in \mathbb{N}$, the state of the channel, $S_t$, is a Bernoulli random variable, i.i.d. across $t$, such that $\Pr(S_t = 1) = \theta_1, \Pr(S_t = 2) = \theta_2$, and $\theta_1 + \theta_2=1$. This can model the two states of a channel, e.g., in quantum illumination \cite{lloyd2008enhanced} where an object's presence ($S_t=1$) or absence ($S_t=2$) affects the behavior of the channel.

\begin{figure*}[htbp]
\center
\begin{tikzpicture}
\fill [black!10] (-7.5,3) rectangle (-3.7,-0.6);

\fill [black!10] (-7.5,-0.9) -- (-0.9,-0.9) -- (-0.9,3) -- (7.5,3) -- (7.5,-3.3) -- (-7.5,-3.3) -- cycle;

\begin{scope}[shift={(-6.8,1)}]
\node (A) at (0,0)[align = center, font=\linespread{1.8}\selectfont] {$A_1$\\$A_2$\\$\vdots$ \\ $A_T$};
\node (Ap) [right=2cm of A, align = center, font=\linespread{1.8}\selectfont] {$A'_1$\\$A'_2$\\$\vdots$ \\ $A'_T$ };
\node at (1.36,1.15)[draw, rectangle, minimum height = 0.5cm, minimum width = 1.1cm, align = center, font=\linespread{1}\selectfont] (E1) {\small $\mathcal{E}^{(W)}_1$};
\node at (1.36,0.38)[draw, rectangle, minimum height = 0.5cm, minimum width = 1.1cm, align = center, font=\linespread{1}\selectfont] (E2) {\small $\mathcal{E}^{(W)}_2$};
\node at (1.36,-0.2){$\vdots$};
\node at (1.36,-1.12)[draw, rectangle, minimum height = 0.5cm, minimum width = 1.1cm, align = center, font=\linespread{1}\selectfont] (ET) {\small $\mathcal{E}^{(W)}_T$};
\draw [color=black, thick, -{Latex[length=1.5mm]}] ($(E1.west)+(-0.5,0)$)--(E1.west);
\draw [color=black, thick, -{Latex[length=1.5mm]}] (E1.east)--($(E1.east)+(0.6,0)$);
\draw [color=black, thick, -{Latex[length=1.5mm]}] ($(E2.west)+(-0.5,0)$)--(E2.west);
\draw [color=black, thick, -{Latex[length=1.5mm]}] (E2.east)--($(E2.east)+(0.6,0)$);
\draw [color=black, thick, -{Latex[length=1.5mm]}] ($(ET.west)+(-0.5,0)$)--(ET.west);
\draw [color=black, thick, -{Latex[length=1.5mm]}] (ET.east)--($(ET.east)+(0.6,0)$);
\end{scope}

\begin{scope} 
\node [right = 2cm of E1, draw, rectangle, minimum height = 0.5cm, minimum width = 1.1cm, align = center, font=\linespread{1}\selectfont] (N1) {\small $\mathcal{N}^{(S_1)}$};
\node [right = 2cm of E2, draw, rectangle, minimum height = 0.5cm, minimum width = 1.1cm, align = center, font=\linespread{1}\selectfont] (N2) {\small $\mathcal{N}^{(S_2)}$};
\node [right = 2cm of ET, draw, rectangle, minimum height = 0.5cm, minimum width = 1.1cm, align = center, font=\linespread{1}\selectfont] (NT) {\small $\mathcal{N}^{(S_T)}$};
\node at (-2.3,0.8){$\vdots$};
\draw [color=black, thick, -{Latex[length=1.5mm]}] ($(N1.west)+(-0.9,0)$)--(N1.west);
\draw [color=black, thick, -{Latex[length=1.5mm]}] (N1.east)--($(N1.east)+(0.9,0)$);
\draw [color=black, thick, -{Latex[length=1.5mm]}] ($(N2.west)+(-0.9,0)$)--(N2.west);
\draw [color=black, thick, -{Latex[length=1.5mm]}] (N2.east)--($(N2.east)+(0.9,0)$);
\draw [color=black, thick, -{Latex[length=1.5mm]}] ($(NT.west)+(-0.9,0)$)--(NT.west);
\draw [color=black, thick, -{Latex[length=1.5mm]}] (NT.east)--($(NT.east)+(0.9,0)$);
\end{scope}

\begin{scope}[shift={(-6.8,-2)}]
	\node[align = center, font=\linespread{1}\selectfont] (B) at (0,0) {$B_1$\\$B_2$\\$\vdots$ \\ $B_T$};
	\draw[color = black, thick, -{Latex[length=1.5mm]}] (0.3,0.7)--(6.7,0.7)--(6.7,3.95)--(7.92,3.95);
	\draw[color = black, thick, -{Latex[length=1.5mm]}] (0.3,0.3)--(6.9,0.3)--(6.9,3.18)--(7.92,3.18);
	\draw[color = black, thick, -{Latex[length=1.5mm]}] (0.3,-0.7)--(7.1,-0.7)--(7.1,1.67)--(7.94,1.67);
	\node at (4.5,-0.1) {$\vdots$};
\end{scope}

\begin{scope}[shift={(-0.75,1)}]
	\node at (0.2,0) [align = center, font=\linespread{1.8}\selectfont] (App) {$A''_1$\\$A''_2$\\$\vdots$ \\ $A''_T$};
	
	\node [right = 2.9cm of N1, draw, rectangle, minimum height = 0.7cm, minimum width = 1cm]  (QI1) {\small $\{\mathcal{M}^{(\hat{s})}\}$};
	\node [right = 2.9cm of N2, draw, rectangle, minimum height = 0.7cm, minimum width = 1cm]  (QI2) {\small $\{\mathcal{M}^{(\hat{s})}\}$};
	\node at (2.5,-0.25) {$\vdots$};
	\node [right = 2.9cm of NT, draw, rectangle, minimum height = 0.7cm, minimum width = 1cm]  (QIT) {\small $\{\mathcal{M}^{(\hat{s})}\}$};
	\draw[color = black, thick, -{Latex[length=1.5mm]}] ($(QI1.west)+(-1.4,0)$) -- (QI1.west);
	\draw[color = black, thick, -{Latex[length=1.5mm]}] ($(QI2.west)+(-1.4,0)$) -- (QI2.west);
	\draw[color = black, thick, -{Latex[length=1.5mm]}] ($(QIT.west)+(-1.4,0)$) -- (QIT.west);

	\node [above right = -0.48cm and 1cm of QI1] (S1) {\small $\hat{S}_1$};
	\node [above right = -0.82cm and 1cm of QI1] (Q1) {\small $Q_1$};
	\node [above right = -0.48cm and 1cm of QI2] (S2) {\small $\hat{S}_2$};
	\node [above right = -0.82cm and 1cm of QI2] (Q2) {\small $Q_2$};
	\node [above right = -0.48cm and 1cm of QIT] (ST) {\small $\hat{S}_T$};
	\node [above right = -0.82cm and 1cm of QIT] (QT) {\small $Q_T$};
	
	\draw[color = black, thick, double] ($(S1.west) + (-1,0)$) -- (S1);
	\draw[color = black, thick, -{Latex[length=1.5mm]}] ($(Q1.west) + (-1,0)$) -- (Q1);
	\draw[color = black, thick, double] ($(S2.west) + (-1,0)$) -- (S2);
	\draw[color = black, thick, -{Latex[length=1.5mm]}] ($(Q2.west) + (-1,0)$) -- (Q2);
	\draw[color = black, thick, double] ($(ST.west) + (-1,0)$) -- (ST);
	\draw[color = black, thick, -{Latex[length=1.5mm]}] ($(QT.west) + (-1,0)$) -- (QT);
	
	\draw[color = black, thick, -{Latex[length=1.5mm]}] ($(S1.east) + (0.04,0)$) -- ($(S1.east) + (1.1,0)$);
	\draw[color = black, thick, -{Latex[length=1.5mm]}] (Q1.east) -- ($(Q1.east) + (1.04,0)$);
	\draw[color = black, thick, -{Latex[length=1.5mm]}] ($(S2.east) + (0.05,0)$) -- ($(S2.east) + (1.1,0)$);
	\draw[color = black, thick, -{Latex[length=1.5mm]}] (Q2.east) -- ($(Q2.east) + (1.04,0)$);
	\draw[color = black, thick, -{Latex[length=1.5mm]}] ($(ST.east) + (-0.02,0)$) -- ($(ST.east) + (1.02,0)$);
	\draw[color = black, thick, -{Latex[length=1.5mm]}] ($(QT.east) + (-0.07,0)$) -- ($(QT.east) + (0.97,0)$);
	
	\node at (5.35,-0.25) {$\vdots$};

	\node at (6.5,0) [draw, rectangle, minimum height = 3cm, minimum width = 1cm, align = center, font=\linespread{1.5}\selectfont] (POVM) {\small POVM \\\small $\big\{\Xi^{(\hat{w})} \big\} $};
	\node[right = 0.5cm of POVM] {$\hat{W}$};
	
	\draw[color = black, thick, -{Latex[length=1.5mm]}] (POVM.east) -- ($(POVM.east)+(0.6,0)$);
\end{scope}

\node at (-6.5,2.8) {\footnotesize \sc Transmitter};
\node at (-2.3,2.8) {\footnotesize \sc Channel};
\node at (6.8,-3) {\footnotesize {\sc Receiver}};
\node at (2.9,-1.5){\small (Instantaneous detection)};
\node at (5.78,-1.5){\small (Decoding)};
\draw[thick, dotted] (-6.3,2.4)--(-6.3,-2.9);
\draw[thick, dotted] (-4.7,2.4)--(-4.7,-2.9);
\draw[thick, dotted] (0.7,2.4)--(0.7,-0.8);
\draw[thick, dotted] (2.9,2.4)--(2.9,-0.8);
\node at (-6.3,-3.1) {$\rho$};
\node at (-4.7,-3.1) {$\sigma$};
\node at (0.7,-1) {$\omega$};
\node at (2.9,-1) {$\kappa$};

\end{tikzpicture}
\caption{Quantum protocol for joint communication and instantaneous detection (JCID).}
\label{fig:protocol}
\end{figure*}
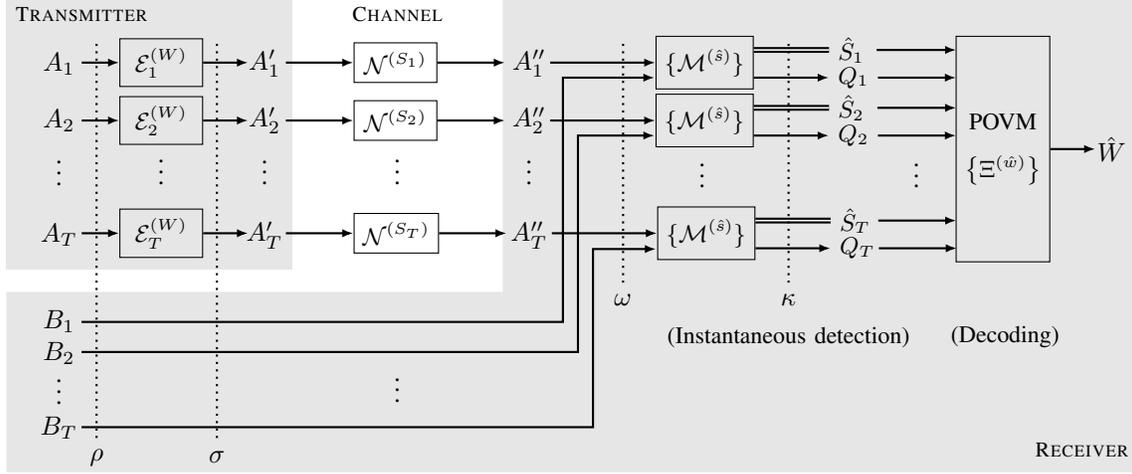

The transmitter and receiver are allowed to share an arbitrarily large compound bipartite quantum system, prepared in advance in a state $\rho$, which can be designed suitably for the desired  performance targets and the channel $\mathcal{N}$ in an offline manner as part of the initial setup prior to the commencement of the JCID operation. Once the JCID operation begins, an independently generated classical message $W$ appears at the transmitter and is encoded into its share of the bipartite quantum system, to be transmitted to the receiver. At each time slot $t\in[T]$ during the JCID operation, the transmitter sends a coded quantum sub-system to the receiver through the channel $\mathcal{N}^{(S_t)}$. The state $S_t$ is not known to the transmitter. The receiver is required to 1) instantly estimate $S_t$ for each time slot $t$, and 2) decode the classical message from the transmitter after  $T$ time slots ($T$ can be chosen arbitrarily large to allow coding for reliable communication across many detection slots). Figure \ref{fig:protocol} illustrates the JCID framework.

Formally, a quantum protocol  $\mathcal{P}$  for JCID is specified by a choice of the following parameters,
{\small
\begin{align} \label{eq:protocol}
	\mathcal{P}\hspace{-0.1cm}=\hspace{-0.1cm}\left(M, T, \rho,   \{\{ \mathcal{E}^{(w)}_t \}_{w\in [M]}\}_{t\in [T]}, \{\mathcal{M}^{(\hat{s})}\}_{\hat{s}\in \{1,2\}},    \{\Xi^{(\hat{w})}  \}_{\hat{w} \in [M]}   \right).
\end{align}
}In order to communicate a classical message $W$ chosen uniformly from $[M]$, over $T$ uses of the quantum channel, the JCID protocol runs as follows. 
Prior to the commencement of the JCID operation, i.e., independent of $W$, quantum systems $A_1B_1 A_2B_2 \cdots A_TB_T$ are distributed in advance to the transmitter and the receiver, such that the systems $\{A_t\}_{t\in [T]}$ are with the transmitter, and the systems $\{B_t\}_{t\in [T]}$ are with the receiver. The dimension of $A_t$ is $a\in \mathbb{N}$ and the dimension of $B_t$ is $b\in \mathbb{N}$, for $t\in [T]$.
The initial state for $A_1B_1 A_2B_2 \cdots A_TB_T$ is set to $\rho \in \mathcal{D}\big((\mathcal{H}_a\otimes \mathcal{H}_b)^{\otimes T}\big)$.
The transmitter is equipped with encoding quantum channels $\{\{\mathcal{E}_t^{(w)}\}_{w\in [M]}\}_{t\in [T]}$, such that each $\mathcal{E}^{(w)}_t \colon \mathcal{D}(\mathcal{H}_a) \mapsto \mathcal{D}(\mathcal{H}_{d})$, which takes  an $a$-dimensional quantum state at the input to  a $d$-dimensional quantum state at the output.

When the JCID operation commences, a random message $W\in [M]$, uniformly and independently generated, appears at the transmitter. The transmitter uses the encoders $\mathcal{E}_t^{(W)}$ over time slots $t\in[T]$, to encode the message into $A_1A_2\cdots A_T$, so that $A_1A_2\cdots A_T$ evolve to $A_1'A_2'\cdots A_T'$. The state of the encoded system $A_1'B_1A_2'B_2\cdots A_T'B_T$ is denoted as $\sigma$. 

At time slot $t\in[T]$, $A_t'$ is sent through the channel $\mathcal{N} = \theta_1\mathcal{N}^{(1)} + \theta_2 \mathcal{N}^{(2)}$ whereby $A_t'$ evolves to $A_t''$. The state for $A_1''B_1 A_2''B_2\cdots A_T''B_T$ is denoted as $\omega$. 
At each time slot $t\in[T]$, the receiver measures $A_t''B_t$ with the quantum instrument $\{\mathcal{M}^{(\hat{s})}\}$, and produces a classical output $\hat{S}_t$ as the instantaneous detection result for that time slot. The remaining quantum system is denoted as $Q_t$. 
The distribution of $\hat{S}_t$ is determined by $P_{\hat{S}_t}(\hat{s}) =  \Tr(\mathcal{M}^{(\hat{s})} (\omega_{A_t''B_t})) $ for $\hat{s} \in \{1,2\}$.

After the $T$ time slots, the receiver collects $\hat{S}_1\cdots \hat{S}_T Q_1\cdots Q_T$, in the state denoted as $\kappa$.
The decoding for the message is done by applying the POVM $\{\Xi^{(\hat{w})}\}$ on $\hat{S}_1\cdots \hat{S}_TQ_1\cdots Q_T$. The measurement result is denoted as $\hat{W}$ and its distribution is determined by $P_{\hat{W}}(\hat{w}) =  \Tr(\Xi^{(\hat{w})} \kappa_{\hat{S}_1\cdots \hat{S}_TQ_1\cdots Q_T})$ for $\hat{w} \in [M]$.

Note that the states mentioned above can be obtained as partial states of the  joint states in \eqref{eq:states} -- \eqref{eq:states_end}.
\begin{figure*}[t]
{\small
	\begin{align} \label{eq:states}
	&\sigma_{WA_1'B_1\cdots A_T'B_T} = \frac{1}{M} \sum_{w\in [M] }  \ket{w}\bra{w}\otimes \big(  \mathcal{E}_1^{(w)} \otimes I_b \otimes \cdots \otimes \mathcal{E}_T^{(w)}\otimes I_b\big)(\rho_{A_1B_1\cdots A_TB_T}), \\
	&\omega_{S_1\cdots S_TWA_1''B_1\cdots A_T''B_T}  =  \sum_{(s_1,\cdots, s_T)\in \{1,2\}^T} \prod_{t=1}^T \theta_{s_t}  \ket{s_1,\cdots,s_T}\bra{s_1,\cdots, s_T}  \notag \\
	&~~~~~~~~~~~~~~~~~~~~~~~~~~~~~~~~~~~~~~~~~~~~~~~~~~~ \otimes I_{M} \otimes \big( \mathcal{N}^{(s_1)} \otimes I_b \otimes \cdots \otimes   \mathcal{N}^{(s_T)} \otimes I_b\big)(\sigma_{WA_1'B_1\cdots A_T'B_T}),  \\
	&\kappa_{\hat{S}_1 \cdots \hat{S}_T Q_1\cdots Q_T} = \sum_{(\hat{s}_1,\cdots, \hat{s}_T) \in \{1,2\}^T} \ket{\hat{s}_1,\cdots, \hat{s}_T}\bra{\hat{s}_1,\cdots, \hat{s}_T} \otimes \big(\mathcal{M} ^{(\hat{s}_1)}\otimes \cdots \otimes \mathcal{M} ^{(\hat{s}_T)}\big) (\omega_{A_1''B_1 \cdots A_T''B_T}). \label{eq:states_end}
\end{align}}
\hrulefill\par
\end{figure*}

\begin{definition}[Unentangled  vs Entangled  Protocols] \label{def:unentangled}
	A quantum protocol for JCID within the described framework is called an unentangled protocol if and only if the initial quantum state is constrained to be separable both in time and in space. Mathematically, this means that $\rho_{A_1B_1A_2B_2\cdots A_TB_T}$ has the form $\sum_{z\in \mathcal{Z}} \lambda_{z} \rho_{A_1\mid z}  \otimes \rho_{B_1\mid z}  \otimes \cdots \otimes \rho_{A_T\mid z}  \otimes \rho_{B_T\mid z} $ for some finite set $\mathcal{Z}$, such that for all $z\in \mathcal{Z}$, $\lambda_{z} \geq 0$,  $\sum_{z\in \mathcal{Z} } \lambda_{z} = 1$, and that $\rho_{A_t \mid z}\in \mathcal{D}(\mathcal{H}_a), \rho_{B_t\mid z}\in \mathcal{D}(\mathcal{H}_b)$ for all $t\in [T]$.  On the other hand, if no such constraint is placed, i.e., the initial state $\rho$ is allowed to be entangled in both time and space, then we refer to the protocol as an Entangled Protocol. 
\end{definition}
Thus, the set of unentangled protocols is a subset of the set of entangled protocols.

\begin{remark}[Common randomness] \label{rem:com_rdm}
The framework  implicitly allows any classical common randomness  to be shared between the transmitter and the receiver in advance.  To see this, note that  $\rho_{A_1B_1\cdots A_TB_T}$ can be set in an arbitrary mixed state, which means each of $A_1, B_1, A_2, B_2,\cdots A_T, B_T$ can contain a classical register that stores a random variable $Z$ that can be encoded by $\mathcal{E}_t, \forall t\in [T]$ and retrieved at the receiver  before measurement. For example, since the protocol is free to choose the initial state, consider a special form of the initial state $\rho_{A_1B_1\cdots A_TB_T}$ such that for $t\in [T]$, $A_t$ is composed of $(\bar{A}_tZ_{A_t})$ and $B_t$ is composed of $(\bar{B}_t Z_{B_t})$, where $Z_{A_t}$ and $Z_{B_t}$ are classical registers that have the value $Z$, which is a classical random variable with distribution $P_Z \in P_{\mathcal{Z}}$ for a finite set $\mathcal{Z}$. Mathematically,
\begin{align}
	&\rho_{A_1B_1\cdots A_TB_T} = \rho_{Z_{A_1}\bar{A}_1 \bar{B}_1Z_{B_1} \cdots Z_{A_T}\bar{A}_T \bar{B}_TZ_{B_T}}  \\
	& =  \sum_{z\in \mathcal{Z}}P_{Z}(z) \bigotimes_{t=1}^T\Big( \ket{z}\bra{z}_{Z_{A_t}}  \otimes \psi_{\bar{A}_t\bar{B}_t} \otimes\ket{z}\bra{z}_{Z_{B_t}}\Big),
\end{align}
for some quantum states $\psi_{\bar{A}_t\bar{B}_t}$ for $t\in [T]$.
\end{remark}

\subsection{Performance metrics}
Given a JCID protocol, let
\begin{align}
	P_{\sf c} \triangleq \max_{w\in [M]} \Pr(\hat{W} \not= W\mid W = w),
\end{align}
be the probability of error for the communication task. 
Let
\begin{align}
	P_{\rm I}^t& \triangleq    \Pr(\hat{S}_t = 2 \mid S_t=1),\\
	P_{\rm II}^t &\triangleq    \Pr(\hat{S}_t = 1 \mid S_t=2),
\end{align}
be the Type I and Type II errors, respectively,  for the instantaneous detection at time slot $t$.

In general, given $\pi_1, \pi_2 \in (0,1)$ such that $\pi_1+\pi_2 = 1$, one is interested in the detection metric $P_{\sf d}^t \triangleq \pi_1 P_{\rm I}^t + \pi_2 P_{\rm II}^t$, which is a  convex combination of $P_{\rm I}^t$ and $P_{\rm II}^t$.\footnote{In the cases $(\pi_1, \pi_2) = (1,0)$ or $(\pi_1, \pi_2) = (0,1)$,  a trivial detection strategy  has $P_{\sf e}=0$, i.e., $\hat{S} = 0$ if $(\pi_1, \pi_2) = (1,0)$, $\hat{S} = 1$ if $(\pi_1, \pi_2) = (0,1)$. We omit these trivial cases as they degenerate to the communication problem.} Define $P_{\sf d} \triangleq \max_{t\in [T]} P_{\sf d}^t$ as the detection performance (for the worst $t$). 
\begin{remark}
	For the detection metric, a Bayesian approach is to consider the average probability of error, which matches the case $\theta_s = \pi_s$ for $s\in \{1,2\}$. However, since the costs for  Type I and Type II errors may be arbitrarily different, in general we consider the whole region of $P_{\rm I}$ and $P_{\rm II}$, which is equivalent to allowing the detection metric to be $\pi_1 P_{\rm I} + \pi_2 P_{\rm II}$ for general $(\pi_1,\pi_2)$. For example, if the two types of error are treated equally, then one should consider $\pi_1=\pi_2 = 0.5$.
\end{remark}

A communication-rate and instantaneous-detection-error\footnote{Unlike the quantum Chernoff bound which is commonly used to gauge  long-term detection performance for a static channel state \cite{shapiro2020quantum},  we consider the probability of detection error as the metric for the instantaneous detection.} pair $(R, P_{\sf e})$ is achievable if and only if for  every  $\epsilon >0$, there exists a protocol  such that
\begin{align}
	\frac{\log_2 M}{T} \geq R-\epsilon, ~~ P_{\sf c} \leq \epsilon, ~~  P_{\sf d} \leq P_{\sf e}.
\end{align}
Define,
{\small
\begin{align}
	\mathfrak{R}_u^* &\triangleq \{(R, P_{\sf e})\colon \mbox{achievable by unentangled protocols}\},\\
	\mathfrak{R}^*_e &\triangleq \{(R, P_{\sf e})\colon \mbox{achievable by entangled protocols}\}.
\end{align}
}

\subsection{Identity-Depolarizing-Erasure (IDE) Channels} \label{sec:special_channel}
Let $\{\ket{i}\}_{i\in[1:d]}$ be the set of computational basis vectors for the input Hilbert space $\mathcal{H}_d$, such that $\ket{i}$ is the $i^{th}$ column of the $d\times d$ identity matrix. The output Hilbert space is $\mathcal{H}_{d+1}$ because we need an extra dimension to correspond to erasures. Let the set of computational basis vectors for $\mathcal{H}_{d+1}$ be  $\{\ket{j}\}_{j\in[1:d+1]}$ where $\ket{j}$ is the $j^{th}$ column of the $(d+1)\times(d+1)$ identity matrix. It will be convenient to denote $\ket{d+1}$ as $\ket{0}$ instead, in order to especially identify erasures at the output.  

We are particularly interested in channels $\mathcal{N}^{(s)}$ that  preserve (identity), depolarize, or erase the input state with probabilities $\alpha_s,\beta_s,\gamma_s$, respectively. In short we refer to this class of channels as Identity-Depolarizing-Erasure (IDE) channels. Specifically, let $\alpha_1, \alpha_2, \beta_1, \beta_2, \gamma_1, \gamma_2 \in [0,1]$, such that $\alpha_s+\beta_s+\gamma_s = 1$ for $s\in \{1,2\}$.   Then $\mathcal{N}^{(s)} \colon \mathcal{D}(\mathcal{H}_d) \mapsto \mathcal{D}(\mathcal{H}_{d+1})$ is a quantum channel such that
\begin{align} \label{eq:def_IDE}
	\mathcal{N}^{(s)}(\rho) = \alpha_s \rho   ~\boxplus~   \beta_s I_d/d   ~\boxplus~   \gamma_s \ket{0} \bra{0}, ~~\forall s \in \{1,2\}.
\end{align}
Note that $\ket{0}\!\bra{0}$ is the $(d+1)\times (d+1)$ matrix with a $1$ in the bottom right corner and zeros elsewhere. Define
\begin{align} \label{eq:def_average_parameters}
	\bar{\alpha} \triangleq \theta_1\alpha_1 + \theta_2 \alpha_2, ~~\bar{\beta} \triangleq \theta_1\beta_1 + \theta_2 \beta_2, ~~ \bar{\gamma} \triangleq \theta_1\gamma_1 + \theta_2 \gamma_2.
\end{align}
Then the overall channel,
\begin{align}
	\mathcal{N}(\rho) = \bar{\alpha} \rho ~\boxplus~ \bar{\beta}I_d/d ~\boxplus~ \bar{\gamma}\ket{0}\bra{0}.
\end{align}

\begin{remark}
	With $\rho$  considered as a $d\times d$ matrix,  we have,
	\begin{align}
		\mathcal{N}^{(s)}(\rho) = \begin{bmatrix}
			\alpha_s\rho+\beta_s I_d/d & 0^{d\times 1} \\ 0^{1\times d} & \gamma_s
		\end{bmatrix}, \forall s\in \{1,2\},
	\end{align}
	and
	\begin{align}
		\mathcal{N}(\rho) = \begin{bmatrix}
			\bar{\alpha} \rho+\bar{\beta} I_d/d & 0^{d\times 1} \\ 0^{1\times d} & \bar{\gamma}
		\end{bmatrix},
	\end{align}
	as $(d+1)\times (d+1)$ matrices.
\end{remark}

\subsection{Useful definitions for IDE Channels}
The following definitions will facilitate compact representations of our results. Recall that IDE channels are specified by parameters $d, (\alpha_s, \beta_s, \gamma_s, \theta_s, \pi_s)_{s\in \{1,2\}}$. Let us also define,
\begin{align}
D&\triangleq \left\{\begin{array}{ll}
d,&\emph{for unentangled settings},\\
d^2,&\emph{for entangled settings},
\end{array}\right.
\end{align}
which intuitively corresponds to the number of signal dimensions accessible to the transmitter in each setting. For example, even if in both cases the transmitter can only access one physical qudit, if that qudit is entangled with another qudit at the receiver, then operations on that one qudit allow the transmitter to manipulate the entangled state which resides in a $d^2$ dimensional space. Intuitively, this is why entangled settings are associated with $D=d^2$ versus $D=d$ for unentangled settings. This expansion of accessible dimensions due to entanglement is also the key to superdense coding schemes \cite{Superdense}.

From the set of all probability mass functions $P_Z\in{P}_{[D]}$ on the domain $[D]$, we will need a small subset, those that take one value, $\Pr(Z=z)=p_1$ for all $z\in [1:n]$ and another value $\Pr(Z=z)=p_2$ for all $z\in[n+1:D]$. Any pmf in this smaller subset is specified by just a $3$-tuple $(n,p_1,p_2)$, and therefore, the set of all such pmf's can be defined as,
\begin{align} \label{eq:def_A_d}
	\mathcal{P}(D) \triangleq \left\{(n,p_1,p_2) \colon \begin{array}{c} n\in [D], p_1,p_2\in [0,1], \\ \mbox{s.t.}~ np_1+(D-n)p_2 = 1 \end{array} \right\}.
\end{align}
Let us  define $(R,P_{\sf e})$ regions $\mathfrak{R}_1(D, \alpha_{[2]}, \beta_{[2]}, \gamma_{[2]}, \theta_{[2]}, \pi_{[2]})$, $\mathfrak{R}_2(D, \alpha_{[2]}, \beta_{[2]}, \gamma_{[2]}, \theta_{[2]}, \pi_{[2]})$ as in \eqref{eq:defR1} and \eqref{eq:defR2}, where 
\begin{align}
\bar{\alpha} &\triangleq \theta_1\alpha_1 + \theta_2 \alpha_2,  \\
\bar{\beta} &\triangleq \theta_1\beta_1 + \theta_2 \beta_2, \\
  \bar{\gamma}& \triangleq \theta_1\gamma_1 + \theta_2 \gamma_2.
  \end{align}
	The key difference between  $\mathfrak{R}_1$ and $\mathfrak{R}_2$ is that $\mathfrak{R}_2$ involves all distributions over the domain $[D]$, i.e., all $P_X\in P_{[D]}$, whereas $\mathfrak{R}_1$ is obtained from $\mathfrak{R}_2$ by restricting the set of distributions from $P_{[D]}$ to the much smaller subset $\mathcal{P}(D)$. Thus, $\mathfrak{R}_1$ is much more efficiently computable than $\mathfrak{R}_2$. On the other hand, the form in $\mathfrak{R}_2$ facilitates the proofs. The fact that the two definitions describe the same region is stated in Theorem \ref{thm:eq} and the proof is presented in Appendix \ref{proof:misc}.

\begin{figure*}[t]
{\small
\begin{align} \label{eq:defR1}
&\mathfrak{R}_1(D, \alpha_{[2]}, \beta_{[2]}, \gamma_{[2]}, \theta_{[2]}, \pi_{[2]}) \triangleq \conv 
\left\{\begin{array}{l} (R,P_{\sf e}) \colon \\ \exists (n,p_1,p_2) \in \mathcal{P}(D)\\
		R \leq -n\big( p_1 \bar{\alpha} + \bar{\beta}/D \big) \log_2\big( p_1 \bar{\alpha} + \bar{\beta}/D \big) -(D-n)\big( p_2 \bar{\alpha} + \bar{\beta}/D \big) \log_2\big( p_2 \bar{\alpha} + \bar{\beta}/D \big) \\ ~~~~~~+ (\bar{\alpha}+\bar{\beta}/D)\log_2 (\bar{\alpha}+\bar{\beta}/D) + (D-1) (\bar{\beta}/D) \log_2 (\bar{\beta}/D)   \\
		P_{\sf e} \geq  n \times \min_{s\in \{1,2\}} \big\{  \pi_s \alpha_s p_1 + \pi_s \beta_s/D  \big\} + (D-n)\times \min_{s\in \{1,2\}} \big\{  \pi_s \alpha_s p_2 + \pi_s \beta_s/D  \big\} \\~~~~~~+ \min\big\{ \pi_1 \gamma_1, \pi_2 \gamma_2 \big\}
		\end{array} \right\}.\\
			& \label{eq:defR2}
		\mathfrak{R}_2(D, \alpha_{[2]}, \beta_{[2]}, \gamma_{[2]}, \theta_{[2]}, \pi_{[2]})
		\triangleq  \conv \left\{ \begin{array}{l} (R,P_{\sf e}) \colon\\
		\exists P_X \in P_{[D]} \\
		R \leq -\sum_{i\in [D]}\big(\bar{\alpha}P_X(i)+\bar{\beta}/D \big)  \log_2 \big(\bar{\alpha}P_X(i)+\bar{\beta}/D \big) \\~~~~~~+ (\bar{\alpha}+\bar{\beta}/D)\log_2 (\bar{\alpha}+\bar{\beta}/D) + (D-1) (\bar{\beta}/D) \log_2 (\bar{\beta}/D)  \\
		P_{\sf e} \geq \sum_{i\in [D]}\min_{s\in \{1,2\}}  \big\{  \pi_s \alpha_s P_X(i) + \pi_s \beta_s/D  \big\} + \min\big\{ \pi_1 \gamma_1, \pi_2 \gamma_2 \big\}
		\end{array} \right\}.
\end{align}}
\hrulefill\par
\end{figure*}

\subsection{Results}
In order to establish an advantage due to quantum entanglement, we need a converse (an impossibility result) for unentangled protocols, and an achievability result for entangled protocols. We start with a converse for unentangled protocols.
\begin{theorem}[General outer bound] \label{thm:une_conv}
{\small
\begin{align}
	\mathfrak{R}^*_u \subseteq \conv \left\{\begin{array}{l} (R,P_{\sf e}) \colon \\ \exists M \in \mathbb{N}, ~~\sigma_1,\sigma_2,\cdots,\sigma_M \in \mathcal{D}(\mathcal{H}_d),  \\~ \sigma \triangleq \frac{1}{M}\sum_{w\in [M]}  \sigma_w, \\ 
	R \leq {\sf H}(\mathcal{N}(\sigma)) - \frac{1}{M}\sum_{w\in [M]} {\sf H}(\mathcal{N}(\sigma_w)), \\ P_{\sf e} \geq \frac{1}{2}\Big( 1-\Vert \pi_1 \mathcal{N}^{(1)}(\sigma) - \pi_2 \mathcal{N}^{(2)}(\sigma) \Vert_1 \Big) \end{array} \right\}.
\end{align}
}
\end{theorem}
\noindent  The proof appears in Section \ref{proof:une_conv}. 
Intuitively, the converse shows that for an unentangled protocol, the utility of quantum systems $B_1,\cdots, B_T$ is limited to enabling classical common randomness between the  transmitter and receiver.

On one hand, the absence of entanglement across time allows the rate $R$ to be bounded by Holevo information without the need for regularization. But on the other hand, as a caveat note that $M$ is unbounded, so the region in Theorem \ref{thm:une_conv} is not directly computable for general channels.  Fortunately, Theorem \ref{thm:une_conv} is still useful in obtaining a computable outer bound for the IDE channels defined in Section \ref{sec:special_channel}. 

The following theorems characterize the performance of unentangled and entangled protocols for IDE channels. 
\begin{theorem} \label{thm:unentangled}
	For IDE channels with unentangled protocols, 
	\begin{align}
	\mathfrak{R}_u^*=\mathfrak{R}_1(d, \alpha_{[2]}, \beta_{[2]}, \gamma_{[2]}, \theta_{[2]}, \pi_{[2]}).
	\end{align}
\end{theorem}
\noindent The converse is based on Theorem \ref{thm:une_conv}, and the achievability follows essentially from a classical strategy with common randomness shared between the transmitter and the receiver.
 
\begin{theorem} \label{thm:superdense_coding}
	For IDE channels with entangled protocols, we have the innerbound 
	\begin{align}
	\mathfrak{R}^*_e\supseteq\underline{\mathfrak{R}^*_e}=\mathfrak{R}_1(d^2, \alpha_{[2]}, \beta_{[2]}, \gamma_{[2]}, \theta_{[2]}, \pi_{[2]}).
	\end{align}
\end{theorem}
\noindent The innerbound is based on a coding scheme that utilizes superdense coding \cite{Superdense}. 
Theorem \ref{thm:unentangled} and Theorem \ref{thm:superdense_coding} are stated in explicitly computable forms to facilitate numerical analysis.

\begin{remark}
	The coding scheme for Theorem \ref{thm:superdense_coding}  uses quantum entanglement across space (between the transmitter and the receiver), but not across time (channel uses). The receiver measures the entangled quantum systems for each channel use in the (general) Bell basis to obtain enough information for the instantaneous detection, and collects the measurements across all channel uses to decode the message. That this relatively simple scheme suffices to achieve significant improvement (Fig. \ref{fig:ex1}) upon the best unentangled scheme especially underscores the utility of entanglement. 
\end{remark}

The proof of $\mathfrak{R}_u^*$ and $\underline{\mathfrak{R}_e^*}$ as presented in Theorem \ref{thm:unentangled} and Theorem \ref{thm:superdense_coding} is provided in Appendix \ref{proof:unentangled}, Appendix \ref{proof:superdense_coding}, respectively, and is based on equivalent alternative (less efficiently computable but more easily provable) forms that are established next.

\begin{theorem}[Equivalent forms]\label{thm:eq}
The regions defined in \eqref{eq:defR1} and \eqref{eq:defR2} are the same, i.e.,
$\mathfrak{R}_1(\cdot) = \mathfrak{R}_2(\cdot)$, where the parameters $(D, \alpha_{[2]}, \beta_{[2]},\gamma_{[2]}, \theta_{[2]}, \pi_{[2]})$ are suppressed as $(\cdot)$. Therefore, Theorems \ref{thm:unentangled} and \ref{thm:superdense_coding} are equivalently stated as
\begin{align}
	\mathfrak{R}_u^* &=\mathfrak{R}_2(d, \alpha_{[2]}, \beta_{[2]}, \gamma_{[2]},\theta_{[2]}, \pi_{[2]}),\label{eq:compact_form_R_u} \\
		 \underline{\mathfrak{R}_e^*} &= \mathfrak{R}_2(d^2, \alpha_{[2]}, \beta_{[2]}, \gamma_{[2]},\theta_{[2]}, \pi_{[2]}). \label{eq:compact_form_R_e}
\end{align}
\end{theorem}
\noindent Theorem \ref{thm:eq} will be proved along with the following observations {\bf O2} -- {\bf O5} in Appendix \ref{proof:misc}.

For insights, let us present some observations that are implied by Theorems \ref{thm:unentangled}, \ref{thm:superdense_coding}, \ref{thm:eq} (see Appendix \ref{proof:misc} for derivations), and some numerical examples. In the following we use $\mathfrak{R}$ to refer to the region $\mathfrak{R}_1$ and $\mathfrak{R}_2$, as they are proved the same.

\begin{figure*}[b]
\center
\begin{tikzpicture} 
\begin{scope}[shift={(0,0)}]
\draw [-{Latex[length=1.5mm]}, thick] (0,0)--(5,0) node [above] {$P_{\sf e}$};
\draw [-{Latex[length=1.5mm]}, thick] (0,0)--(0,3) node [right] {$R$};
\draw [thick] (1,0)--(1.25,1)--(2,1.75)--(3,2)--(4.5,2);
\draw[pattern={dots}] (1,0)--(1.25,1)--(2,1.75)--(3,2)--(4.2,2) --(4.2,0) -- cycle;
\draw[thick, dashed] (0,2)--(3,2);
\draw[thick, dashed] (3,0)--(3,2);
\node at (-0.2,0) {\small $O$};
\node at (1,-0.2) {\small $P_{{\sf e}, \min}$};
\node at (3,-0.2) {\small $P_{{\sf e}, *}$};
\node at (-0.5, 2) {\small $R_{\max}$};
\node at (2.5,-0.75) {\small Trade-off between $R$ and $P_{\sf e}$};
\end{scope}

\begin{scope}[shift={(8,0)}]
\draw [-{Latex[length=1.5mm]}, thick] (0,0)--(5,0) node [above] {$P_{\sf e}$};
\draw [-{Latex[length=1.5mm]}, thick] (0,0)--(0,3) node [right] {$R$};
\draw [thick] (1,0)--(1,2)--(4.5,2);
\draw[pattern={dots}] (1,0)--(1,2)--(4.2,2)--(4.2,0);

\draw[thick, dashed] (0,2)--(1,2);
\node at (-0.2,0) {\small $O$};
\node at (1.35,-0.2) {\small $P_{{\sf e}, \min}$ ($P_{{\sf e}, *}$)};

\node at (-0.5, 2) {\small $R_{\max}$};
\node at (2.5,-0.75) {\small No trade-off between $R$ and $P_{\sf e}$};
\end{scope}
\end{tikzpicture}
\caption{Two typical $\mathfrak{R}(\cdot)$ regions are shown (dotted areas). For the case on the LHS, there is an interval of $P_e$ in which (the highest) $R$ is strictly increasing with respect to $P_{\sf e}$. We say that there is a trade-off between $R$ and $P_{\sf e}$ within this interval. For the case shown on the RHS, there is no such trade-off.} \label{fig:region}
\end{figure*}
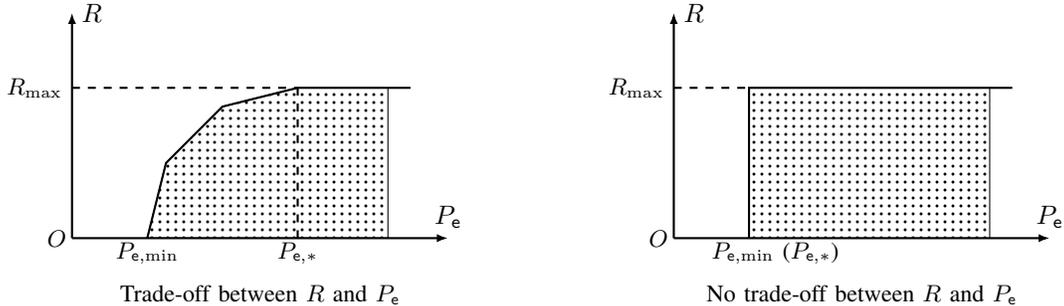

\begin{enumerate}[start=1,label={\bfseries O\arabic*:},wide = 3pt, leftmargin = 0em]
\item  Typical $\mathfrak{R}(\cdot)$ regions, with and without $(R,P_{\sf e})$ tradeoffs, are as illustrated in Fig. \ref{fig:region}.
\item Closed form expression for the maximum rate can be obtained as, 
$R_{\max} = -(\bar{\alpha}+\bar{\beta})\log_2[(\bar{\alpha}+\bar{\beta})/D]+(\bar{\alpha}+\bar{\beta}/D)\log_2(\bar{\alpha}+\bar{\beta}/D)+\frac{D-1}{D}\bar{\beta}\log_2 (\bar{\beta}/D)$. 
\item Closed form expression for the minimum probability of detection error is $P_{{\sf e}, \min} = \min\{ \pi_1 (\alpha_1+\beta_1/D), \pi_2 (\alpha_2+\beta_2/D) \} + \frac{D-1}{D}   \min\{ \pi_1 \beta_1, \pi_2 \beta_2 \} + \min\{ \pi_1 \gamma_1, \pi_2 \gamma_2 \}$.
\item $R_{\max}$ is attainable as long as $P_{\sf e} \geq \min\{\pi_1(\alpha_1+\beta_1), \pi_2(\alpha_2+\beta_2)\}+\min\{\pi_1\gamma_1, \pi_2 \gamma_2\} \triangleq P_{{\sf e}, *}$.
	\item Each of the following conditions is sufficient for $\mathfrak{R}(\cdot)$ to have no trade-off between $R$ and $P_{\sf e}$. 
	\begin{enumerate}[label=\arabic*), align =left]
	\item $\pi_1\alpha_1= \pi_2\alpha_2$.
	\item ($\pi_1\alpha_1> \pi_2\alpha_2$) and ($p_{th}\leq 0$ or $p_{th}\geq 1$), where $p_{th}\triangleq \frac{\pi_2 \beta_2 - \pi_1 \beta_1}{(\pi_1\alpha_1 -\pi_2 \alpha_2)D}$.
	\end{enumerate}
\end{enumerate}

\begin{example}\label{ex1}
	Let $d = 16$, $(\alpha_1, \beta_1, \gamma_1)=(1,0,0)$, $(\alpha_2, \beta_2, \gamma_2)=(0,1,0)$, $\pi_1 = \pi_2 = 0.5$. The physical assumption for this setting is that when an obstacle (e.g., detection target) is absent, the channel is noiseless, and when the obstacle is present, it totally breaks the channel and the receiver sees a uniformly random output. Fig. \ref{fig:ex1} illustrates $\mathfrak{R}^*_u$ and $\underline{\mathfrak{R}^*_e}$ for $(\theta_1,\theta_2) \in \{(0.01,0.99), (0.02,0.98), (0.05,0.95)\}$. For these cases, a larger $\theta_1$ implies a better communication quality of the overall channel $\mathcal{N}$. For unentangled protocols, the communication rate is non-zero if and only if $P_{\sf e} \geq 1/32 \approx 0.0312$, whereas entangled protocols achieve a non-zero  communication rate by superdense coding if $P_{\sf e} \geq 1/512 \approx 0.002$. The ratio between the two thresholds is $d$, which resembles the signal-to-ratio enhancement by entanglement observed in \cite{lloyd2008enhanced}. At $P_{\sf e} = 0.0312$, for $\theta_1 = 0.01,0.02,0.05$, entangled protocols already achieve a rate approximately  $9.6,7.5,5.0$ times (respectively) the optimal rate of unentangled protocols at $P_{\sf e} = 0.5$.  This rate gain factor is larger when $\theta_1$ is smaller, i.e., when $\mathcal{N}$ is worse for communication. It can be analytically shown (similar to  \cite{bennett_shor_capacity}) that the limit of the rate gain at $P_{\sf e}=0.5$ for entangled vs unentangled protocols is $d+1$ as $\theta_1 \to 0$. 
\begin{figure}[htbp!]
\center
	\includegraphics[width=0.5\textwidth]{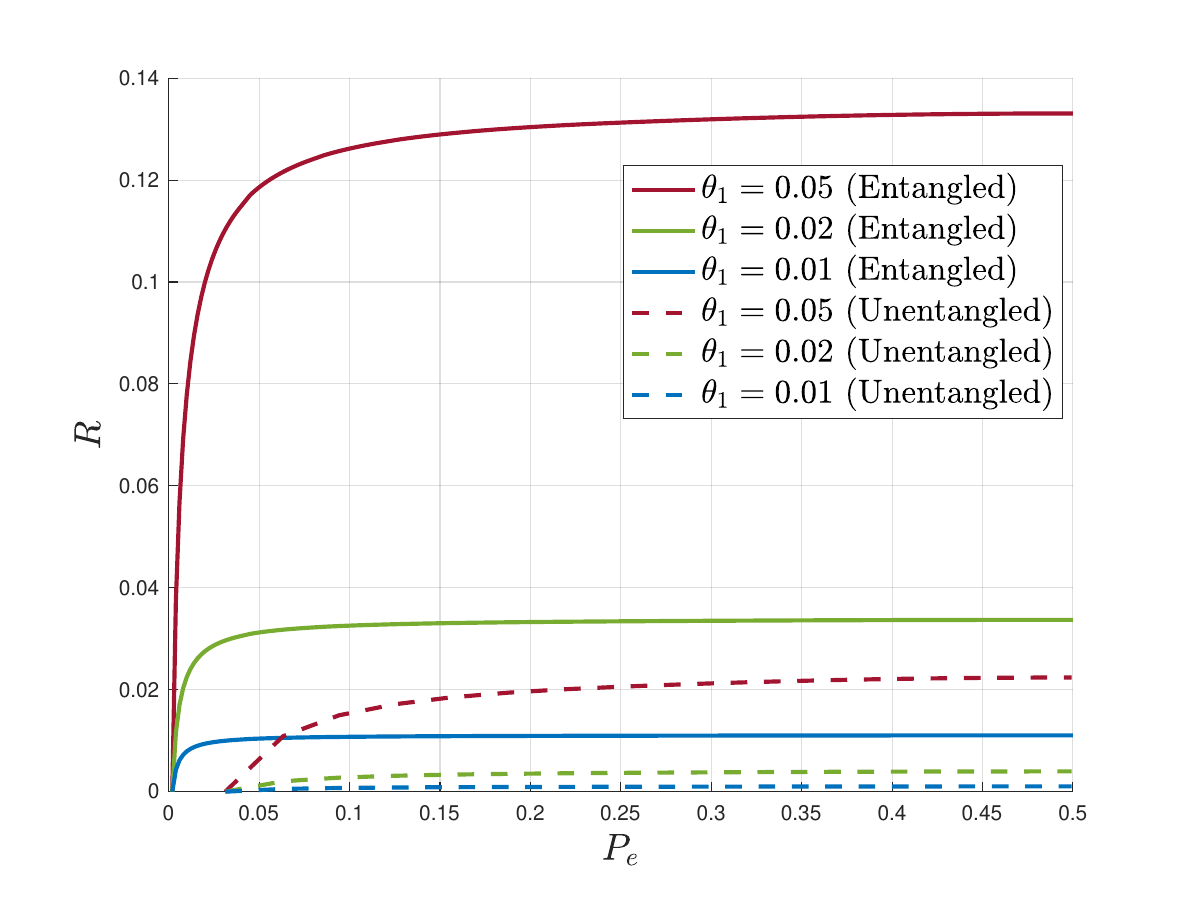}
\caption{Corresponding to Example \ref{ex1}, highest $R$ is shown  with respect to $P_{\sf e}$ for $\mathfrak{R}^*_u$ (tight for unentangled protocols, dashed curves) and for $\underline{\mathfrak{R}^*_e}$ (achievable for entangled protocols, solid curves). We note that for each $\theta_1$, the proposed entangled protocol achieves a higher rate than the corresponding capacity of unentangled protocols, for any detection error probability.} \label{fig:ex1}
\end{figure}
\end{example}

\begin{example}\label{ex2}
	Let $d= 16$, $\theta_1 = \theta_2 = 0.5$, $(\alpha_1, \beta_1, \gamma_1) = (0.8,0.1,0.1)$, $\pi_1 = \pi_2 = 0.5$.  Fig. \ref{fig:ex2} illustrates $\mathfrak{R}^*_u$  and $\underline{\mathfrak{R}^*_e}$ for $(\alpha_2, \beta_2, \gamma_2) \in \{(0.2,0.7,0.1),(0.4,0.5,0.1),(0.6,0.3,0.1),(0.8,0.1,0.1)\}$. For these cases, a larger $\alpha_2$ implies a better communication quality of the overall channel $\mathcal{N}$.  Note that as $\alpha_2$  becomes closer to  $\alpha_1=0.8$,  the two channel states become more similar, which  hurts the detection performance, whereas a larger $\alpha_2$ helps the communication performance because the overall channel  is better for communication.  
 
\begin{figure}[htbp!]
\center
	\includegraphics[width=0.5\textwidth]{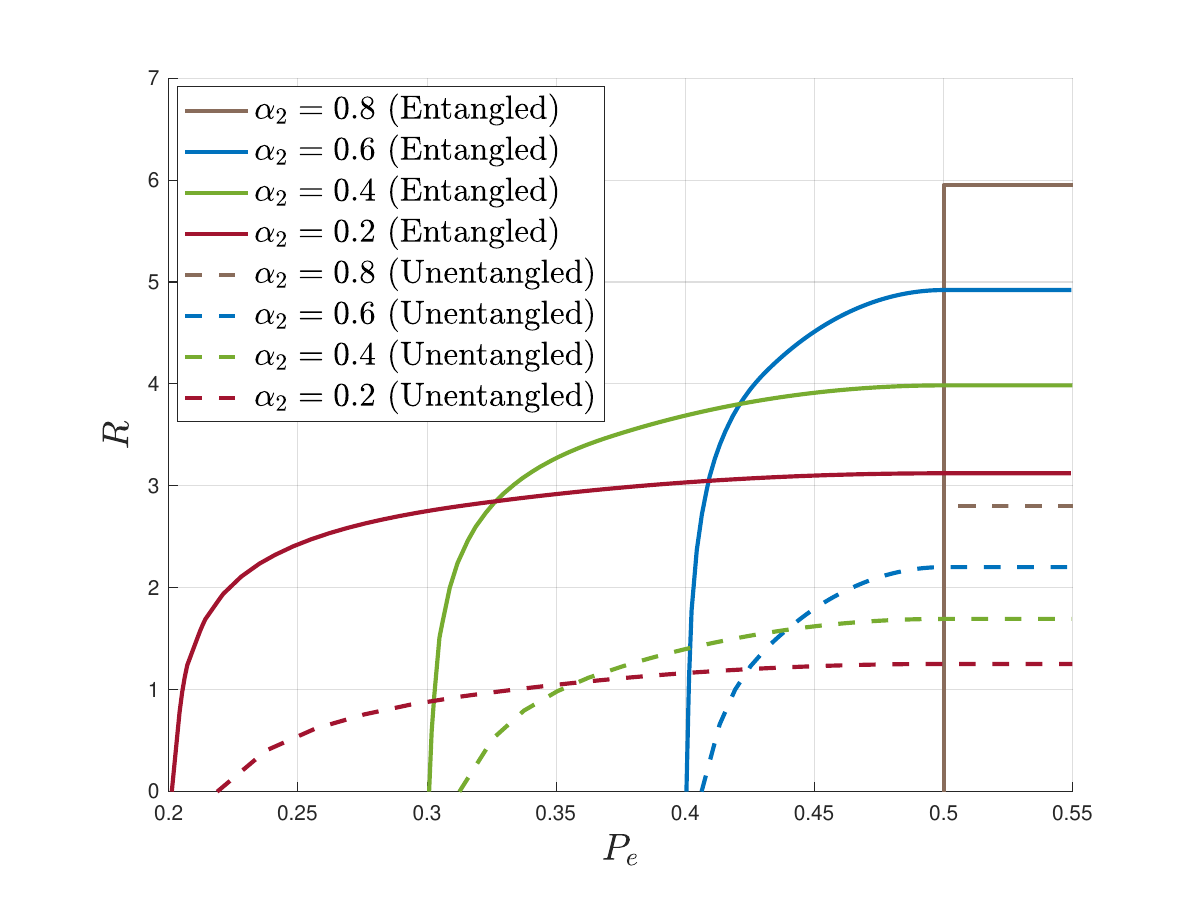}
\caption{Corresponding to Example \ref{ex2}, highest $R$ is shown with respect to $P_{\sf e}$ for $\mathfrak{R}^*_u$  and $\underline{\mathfrak{R}^*_e}$. }\label{fig:ex2}
\end{figure}
\end{example}

\section{JCID with unreliable entangled resource}
In the JCID framework, we assume the systems $B_1,B_2,\cdots, B_T$ keep intact during the evolution of the systems $A_1,A_2,\cdots, A_T$. Indeed, in deriving the achievability result of Theorem \ref{thm:superdense_coding},  the superdense coding protocol is invoked, which requires the transmitter and the receiver to share a batch of maximally entangled states ($d$-ary Bell states). 
Practically, entanglement is a vulnerable resource and may be hard to preserve. To study how the protocol performs when the entanglement is unreliable, we consider the following model building upon the JCID framework. 
\begin{figure}[htbp]
\center
\begin{tikzpicture}
	\node (B1) at (0,0) {$\bar{B}_1$};
	\node (B2)  [below = 0.1cm of B1] {$\bar{B}_2$};
	\node [below = -0.2cm of B2] {$\vdots$};
	\node (BT) [below = 0.6cm of B2]  {$\bar{B}_T$};
	\node (N1) [right = 1cm of B1, draw, rectangle, minimum width = 1cm, minimum height = 0.3cm]  {$\mathcal{N}_B$};
	\node (N2) [right = 1cm of B2, draw, rectangle, minimum width = 1cm, minimum height = 0.3cm]  {$\mathcal{N}_B$};
	\node [below = -0.2cm of N2] {$\vdots$};
	\node (NT) [right = 1cm of BT, draw, rectangle, minimum width = 1cm, minimum height = 0.3cm]  {$\mathcal{N}_B$};
	\draw[-{Latex[length=1.5mm]}, thick ] (B1.east) -- (N1.west);
	\draw[-{Latex[length=1.5mm]}, thick ] (N1.east) -- ($(N1.east)+(1,0)$);
	\draw[-{Latex[length=1.5mm]}, thick ] (B2.east) -- (N2.west);
	\draw[-{Latex[length=1.5mm]}, thick ] (N2.east) -- ($(N2.east)+(1,0)$);
	\draw[-{Latex[length=1.5mm]}, thick ] (BT.east) -- (NT.west);
	\draw[-{Latex[length=1.5mm]}, thick ] (NT.east) -- ($(NT.east)+(1,0)$);
	\node at (4.75,-1) {\small (To measurements)};
	
	\node (Z) [above = 0.3cm of B1] {$Z$};
	\draw[-{Latex[length=1.5mm]}, thick ] (Z.east) -- ($(Z.east)+(3.09,0)$);
	\draw[-{Latex[length=1.5mm]}, thick ] (Z.west) -- ($(Z.west)+(-1.5,0)$);
	\node at (0,1.2) {\small (Common randomness)};
	\node at (-1.5,0.5) {\small (To Transmitter)};
\end{tikzpicture}
\caption{Modeling unreliable entangled resource by passing receiver-side entangled quantum resources $\bar{B}_1,\bar{B}_2,\cdots, \bar{B}_T$  through (independent) depolarizing channels $\mathcal{N}_B$.}
\label{fig:unreliable_entanglement}
\end{figure}
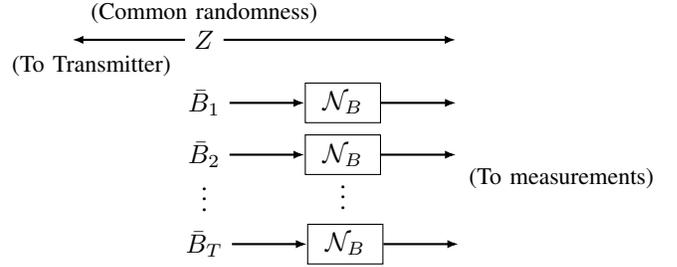
Specifically, we assume that the transmitter and the receiver can share unlimited common randomness (as is also allowed by the JCID protocol, see Remark \ref{rem:com_rdm}), and we model the unreliable entanglement as depolarizing channels  applied to $\bar{B}_1, \bar{B}_2,\cdots, \bar{B}_T$, with a certain probability. See Fig. \ref{fig:unreliable_entanglement} for an explanation. Mathematically,
\begin{align}
	\mathcal{N}_B(\rho) = \widetilde{\alpha} \rho + \widetilde{\beta}I_b/b,
\end{align}
where $\widetilde{\alpha}, \widetilde{\beta} $ are coefficients in $[0,1]$ and $\widetilde{\alpha} + \widetilde{\beta} =1$.  The framework does not pose constraints on the dimension of $B_t$'s, i.e., $b$ can be arbitrarily large. Denote $\mathfrak{R}_{ue}^{*}$ as the region of all achievable $(R,P_{\sf e})$ tuples in the framework with unreliable entangled resource.

\subsection{Result}
Define
\begin{align}
	\ddot{\alpha}_s \triangleq \alpha_s \widetilde{\alpha}, ~~~ \ddot{\beta}_s \triangleq \alpha_s \widetilde{\beta}+ \beta_s,   ~~~ \forall s\in \{1,2\},
\end{align}
and
\begin{align}
	\bar{\ddot{\alpha}} \triangleq \sum_{s\in \{1,2\}}\theta_s \ddot{\alpha}_s  , ~~~ \bar{\ddot{\beta}} \triangleq \sum_{s\in \{1,2\}}\theta_s \ddot{\beta}_s.
\end{align}
It can be verified that $\ddot{\alpha}_s + \ddot{\beta}_s +  \gamma_s = 1$ for $s\in \{1,2\}$ and that $\bar{\ddot{\alpha}}+\bar{\ddot{\beta}} +  \bar{\gamma}  = 1$. We are  able to establish the following innerbound on $\mathfrak{R}_{ue}^{*}$ with respect to $d, \{\ddot{\alpha}_s, \ddot{\beta}_s, \ddot{\gamma}_s, \theta_s, \pi_s\}_{s\in \{1,2\}}$. 
\begin{theorem}\label{thm:unreliable}
$\mathfrak{R}_{ue}^{*} \supseteq \conv ( \mathfrak{R}_{sdc} \cup \mathfrak{R}_u^*)$,
where  
\begin{align}
	\mathfrak{R}_{sdc} = \mathfrak{R}_1(d^2, \ddot{\alpha}_{[2]}, \ddot{\beta}_{[2]}, \gamma_{[2]},  \theta_{[2]}, \pi_{[2]}).\label{eq:region_unreliable_compact}
\end{align}
with $\mathfrak{R}_1(\cdot)$ defined as  in \eqref{eq:defR1}.
\end{theorem}
\noindent The proof is given in Section \ref{proof:superdense_coding}, based on the equivalent form $\mathfrak{R}_2$.

\begin{example}\label{ex3}
	Let $d = 16$, $(\alpha_1, \beta_1, \gamma_1)=(1,0,0)$, $(\alpha_2, \beta_2, \gamma_2)=(0,1,0)$, $\pi_1 = \pi_2 = 0.5$. These parameters are the same as Example 1. For the rest of the parameters, we compare the cases for $(\widetilde{\alpha}, \widetilde{\beta}) = (0.95,0.05)$, $(\widetilde{\alpha}, \widetilde{\beta}) = (0.8,0.2)$, $(\widetilde{\alpha}, \widetilde{\beta}) = (0.5,0.5)$, i.e., with decreasingly reliable entanglement, for $(\theta_1, \theta_2) \in \{(0.95,0.05), (0.5,0.5)\}$ as shown in Fig. \ref{fig:ex3}.
Note that the achievable region obtained with unreliable entanglement can still be significantly larger than that obtained with unentangled protocols for a wide range of $\widetilde{\alpha}$ (which models the reliability of entanglement). 
	This shows that unreliable entanglement is still useful in improving the performance of communication and instantaneous detection. 
	Unreliable entanglement can have a stronger impact on $P_{{\sf e}, \min}$ than on $R_{\max}$. For example, for $\theta_1= 0.05$ (LHS), $P_{{\sf e}, \min}$  at $\widetilde{\alpha} = 0.95$ is already very close to the $P_{{\sf e}, \min}$ for unentangled protocols, whereas $R_{\max}$  at $\widetilde{\alpha} = 0.95$ is still close to the $R_{\max}$ for protocols with reliable entanglement.
	The comparison between the LHS and RHS shows that the gain from unreliable entangled resource may be less significant when the channel is better for communication. For example, in RHS (good channel), the region with unreliable entanglement at $\widetilde{\alpha} = 0.5$ is very close to the region with unentangled protocols, whereas in LHS (bad channel), the two regions are still mostly distinguishable (for a large range of $P_{\sf e}$).

\begin{figure*}[htbp]
    \center
    \begin{subfigure}[t]{0.48\textwidth}
        \center 
        \includegraphics[width= \textwidth]{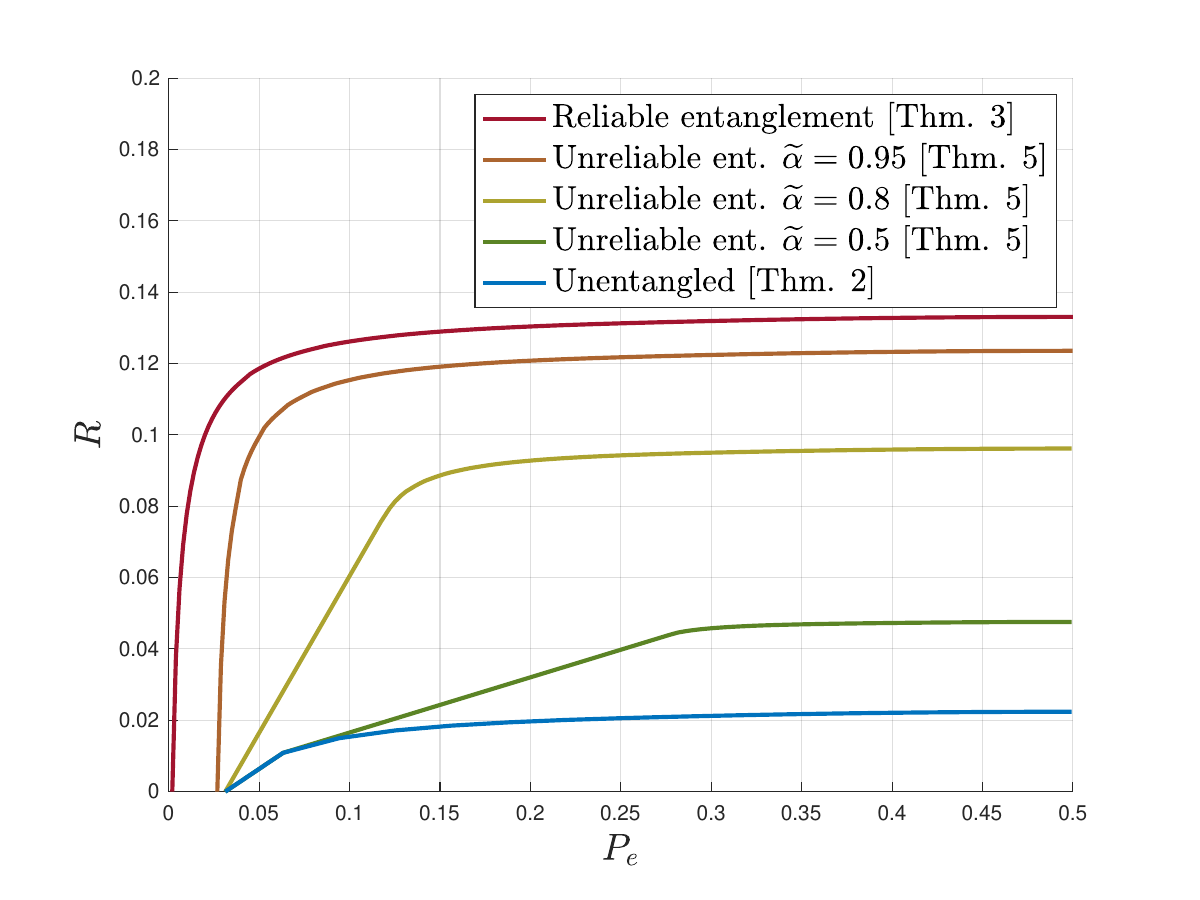}
        \caption{$(\theta_1,\theta_2) = (0.05,0.95)$}
    \end{subfigure} 
    \bigskip%
    \begin{subfigure}[t]{0.48\textwidth}
        \center
        \includegraphics[width= \textwidth]{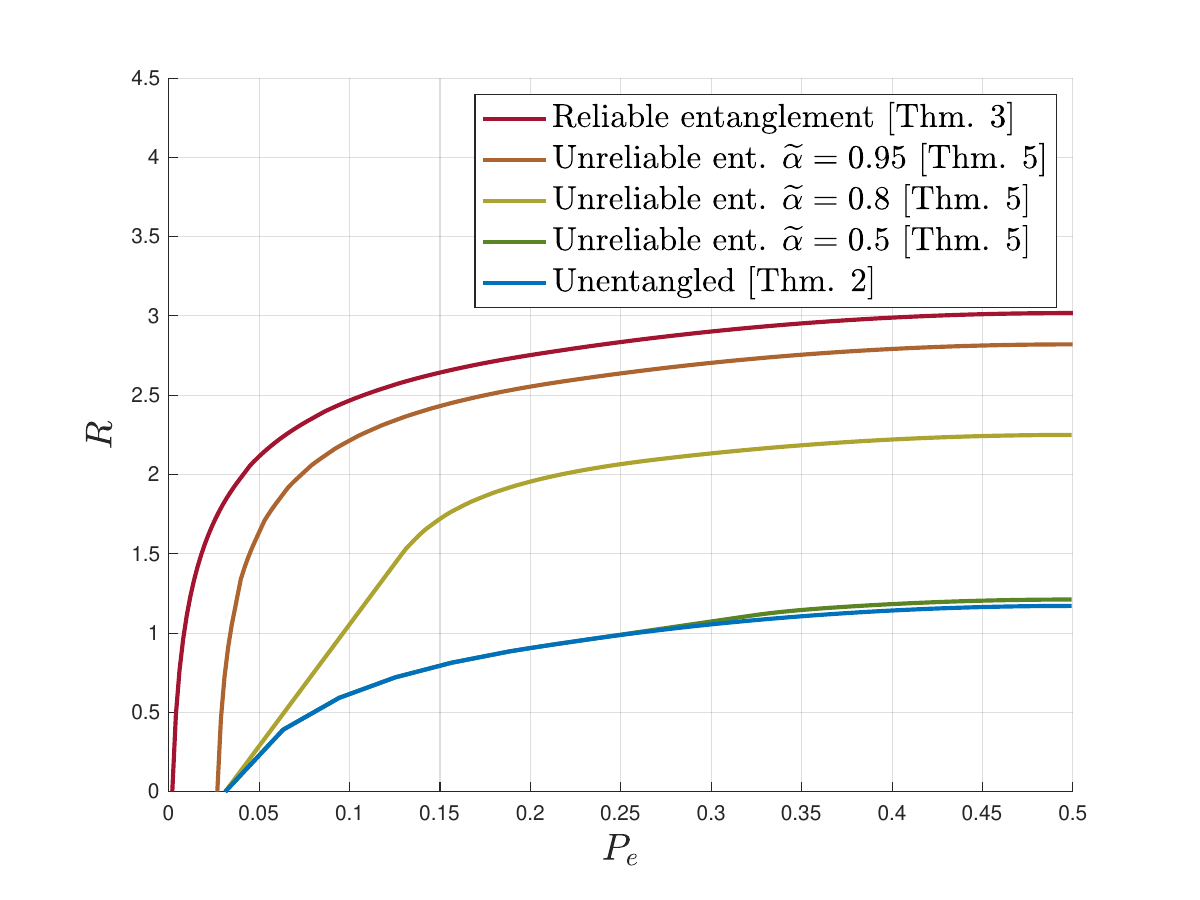}
        \caption{$(\theta_1,\theta_2) = (0.5,0.5)$}
    \end{subfigure}
    \caption{Corresponding to Example \ref{ex3}, highest $R$ is shown with respect to $P_{\sf e}$ for $\underline{\mathfrak{R}_{e}^*}$ (Theorem \ref{thm:superdense_coding}), $\conv(\mathfrak{R}_{sdc}\cup \mathfrak{R}_{u}^*))$ (Theorem \ref{thm:unreliable}) and $\mathfrak{R}^*_u$ (Theorem \ref{thm:unentangled}).  } \label{fig:ex3}
\end{figure*}
\end{example}

\section{Conclusion}
 The finding of significant performance gain from quantum entanglement simultaneously for both communication and instantaneous detection is an encouraging outcome, especially since the gain is accessible through relatively simple entangled protocols over the \emph{best} unentangled scheme. Generalizations, e.g., to non-binary, non i.i.d. channel states, and two-phase coding schemes, are likely to be  challenging but are well-motivated by this work. This work also motivates future efforts towards understanding the utility of quantum entanglement for generalized communication, detection and estimation tasks, especially in distributed/multiuser/network settings  \cite{Yao_Jafar_Sum_MAC} where the utility of entanglement could potentially be further amplified.

\appendices
\section{Proofs of {\bf O2} -- {\bf O5} and Theorem \ref{thm:eq}} \label{proof:misc}
In the following, given any $P_X \in P_{[D]}$, define the compact notation $P_X(i) \triangleq p_i$ for $i\in [D]$, and ${\bm  p} \triangleq (p_1,p_2,\cdots, p_D)$.
Let us recall the definition of $\mathfrak{R}_2$ from \eqref{eq:defR2} into \eqref{eq:def_Rd_recall}, wherein we define $\overline{R}({\bm p})$ and $\underline{P_{\sf e}}({\bm p})$  to represent the bounds on $R$ and $P_{\sf e}$, respectively, given a distribution ${\bm p} \in P_{[D]}$.
\begin{figure*}[b]
\hrulefill\par
{\small
\begin{align} \label{eq:def_Rd_recall}
	\mathfrak{R}_2(\cdot) \triangleq \conv \left\{  \begin{array}{l} (R,P_{\sf e})\colon \\ \exists {\bm p}=(p_1,p_2,\cdots, p_D) \in P_{[D]} \\
		R \leq -\sum_{i\in [D]}\big(\bar{\alpha}p_i +\bar{\beta}/D \big)  \log_2 \big(\bar{\alpha}p_i +\bar{\beta}/D \big)  \\ ~~~~+ (\bar{\alpha}+\bar{\beta}/D) \log_2 (\bar{\alpha}+\bar{\beta}/D) + (D-1) (\bar{\beta}/D) \log_2 (\bar{\beta}/D) \hspace{0.6cm} \triangleq ~\overline{R}({\bm p}) \\
		P_{\sf e} \geq \sum_{i\in [D]}\min_{s\in \{1,2\}}  \big\{  \pi_s \alpha_s p_i  + \pi_s \beta_s/D  \big\} + \min\big\{ \pi_1 \gamma_1, \pi_2 \gamma_2 \big\} \triangleq ~  \underline{P_{\sf e}}({\bm p})
		\end{array} \right\}.
\end{align}}
\end{figure*}

\subsection{Proof of {\bf  O2}} 
For $k\in [D]$, let ${\bm p}^{(k)} = (p_{k}, p_{k+1}, \cdots, p_D,p_1, \cdots, p_{k-1})$ be a cyclic permutation of ${\bm p}$. Note that ${\bm p} = {\bm p}^{(1)}$. Due to the symmetry in the definition of $\overline{R}({\bm p})$, 
$\overline{R}({\bm  p}^{(k)}) = \overline{R}({\bm  p})$ for $k\in [D]$. 
It can be verified that $\overline{R}({\bm  p})$ is concave with respect to ${\bm p}$. It follows that
\begin{align}
	\overline{R}([1/D,1/D,\cdots, 1/D]) = \overline{R}\Big( (1/D) \sum_{k\in [D]} {\bm p}^{(k)} \Big) \notag \\
	\geq (1/D) \sum_{k\in [D]} \overline{R}({\bm p}^{(k)}) = \overline{R}({\bm p}).
\end{align}
Therefore, the bound $\overline{R}({\bm p})$ for $R$ is maximized when ${\bm p} = [1/D,1/D,\cdots, 1/D]$. This yields
\begin{align}
	&R_{\max} = -(\bar{\alpha}+\bar{\beta})\log_2[(\bar{\alpha}+\bar{\beta})/D] \notag \\
	&+(\bar{\alpha}+\bar{\beta}/D)\log_2(\bar{\alpha}+\bar{\beta}/D)+\frac{D-1}{D}\bar{\beta}\log_2 (\bar{\beta}/D),
\end{align}
which proves {\bf O2}.   \hfill \qed

\subsection{Proof of {\bf O3}}
The bound for $P_{\sf e}$ is
{\small
\begin{align}
	 \underline{P_{\sf e}}({\bm p})  &= \sum_{i\in [D]} \min \big\{ \underbrace{\pi_1 \alpha_1 p_i + \pi_1 \beta_1/D}_{f_1(p_i)} , ~ \underbrace{\pi_2 \alpha_2 p_i + \pi_2 \beta_2/D}_{f_2(p_i)} \big\} \notag \\
	& ~~~~~~~~~~~~+ \min\big\{ \pi_1 \gamma_1, \pi_2 \gamma_2 \big\}  \label{eq:Pe_bound} \\
	&= \frac{1}{2}\sum_{i\in [D]} \big(f_1(p_i)+f_2(p_i)\big) + \min\big\{ \pi_1 \gamma_1, \pi_2 \gamma_2 \big\} \notag\\
	&~~~~~~~~~~~~- \frac{1}{2}\sum_{i\in [D]} \vert f_1(p_i)-f_2(p_i) \vert \label{eq:use_min_abs} \\
	&= \frac{1}{2} \big( \pi_1(\alpha_1+\beta_1)+\pi_2(\alpha_2+\beta_2) \big)+ \min\big\{ \pi_1 \gamma_1, \pi_2 \gamma_2 \big\} \notag \\
	&~~~~~~~~~~~- \frac{1}{2} \underbrace{\sum_{i\in [D]} \vert f_1(p_i)-f_2(p_i) \vert}_{F({\bm p})}. \label{eq:def_F_function}
\end{align}
}In \eqref{eq:Pe_bound}, we define the functions $f_s(p) \triangleq \pi_s\alpha_s p + \pi_s \beta_s/D$ for $s\in \{1,2\}$. Step \eqref{eq:use_min_abs} is because $\min\{u, v\}=(1/2)(u+v-|u-v|)$ for reals $u,v$.  In \eqref{eq:def_F_function}, we define the function $F({\bm p})$. For $i \in [D]$, let ${\bm e}^{(i)}$ be the $i^{th}$ row of the identity matrix $I_d$. Then due to symmetry $F({\bm e}^{(i)}) = F({\bm e}^{(1)})$ for $i\in [D]$.
Since sum of convex functions is convex, we obtain that $F({\bm p})$ is convex with respect to ${\bm p}$. Therefore,
{\small
\begin{align}
	F({\bm p}) = F\Big( \sum_{i\in [D]} p_i {\bm e}^{(i)} \Big) \leq \sum_{i\in [D]}p_i F({\bm e}^{(i)}) = F({\bm e}^{(1)}).
\end{align}
}It follows from \eqref{eq:def_F_function} that
{\small
\begin{align}
	\underline{P_{\sf e}}({\bm p}) &\geq \frac{1}{2} \big( \pi_1(\alpha_1+\beta_1)+\pi_2(\alpha_2+\beta_2) \big) + \min\big\{ \pi_1 \gamma_1, \pi_2 \gamma_2 \big\} \notag \\
	&~~~~~~~~~~~~~~~~~~~~~~- \frac{1}{2}F({\bm e}^{(1)}) \\
	&= \min\{ \pi_1 (\alpha_1+\beta_1/D), \pi_2 (\alpha_2+\beta_2/D) \} \notag\\
	&~~+ \frac{D-1}{D}   \min\{ \pi_1 \beta_1, \pi_2 \beta_2 \} + \min\{ \pi_1 \gamma_1, \pi_2 \gamma_2 \} \\
	&= P_{{\sf e}, \min},
\end{align}
}and the inequality is saturated by letting ${\bm p} = {\bm e}^{(1)}$. This proves {\bf O3}. \hfill \qed

\subsection{Proof of {\bf O4}}
Plugging in ${\bm p} = [1/D,1/D,\cdots, 1/D]$, we obtain that
{\small
\begin{align}
	& \overline{R}({\bm p}) = R_{\max}, \\
	& \underline{P_{\sf e}}({\bm p}) = \min\{\pi_1(\alpha_1+\beta_1), \pi_2(\alpha_2+\beta_2)\}+\min\{\pi_1\gamma_1, \pi_2 \gamma_2\}.
\end{align}
}Therefore, as long as $P_{\sf e} \geq \min\{\pi_1(\alpha_1+\beta_1), \pi_2(\alpha_2+\beta_2)\}+\min\{\pi_1\gamma_1, \pi_2 \gamma_2\}$, $R = R_{\max}$ is attainable in $\mathfrak{R}_2(\cdot)$. This proves {\bf O4}. \hfill \qed

\subsection{Proof of {\bf O5}}
In \eqref{eq:Pe_bound}, note that $f_1(p)$ and $f_2(p)$ are monotonic linear functions with respect to $p \in [0,1]$. For the special case of $\pi_1 \alpha_1 = \pi_2 \alpha_2$,  we observe that $\underline{P_{\sf e}}({\bm p})$ is independent of ${\bm p}$ (because $\sum_{i}p_i = 1$), which makes $\underline{P_{\sf e}}({\bm p})$ a constant. Recall that $\overline{R}({\bm p})$ is maximized to $R_{\max}$ for ${\bm p} = [1/D,1/D,\cdots, 1/D]$. This means there is no trade-off between $R$ and $P_{\sf e}$ in $\mathfrak{R}_2(\cdot)$.
Next consider $\pi_1 \alpha_1  \not= \pi_2 \alpha_2$. Due to symmetry, without loss of generality let us say $\pi_1 \alpha_1  > \pi_2 \alpha_2$.
Define  a threshold value,
\begin{align}
	p_{th}  \triangleq \frac{\pi_2 \beta_2 - \pi_1 \beta_1}{(\pi_1\alpha_1 -\pi_2 \alpha_2)D}.
\end{align}
It follows that,
\begin{align} \label{eq:min_f1_f2}
		\min \{ f_1(p_i), f_2(p_i)\} =  \begin{cases}
			f_1(p_i), & \mbox{if}~ p_i \leq p_{th} \\
			 f_2(p_i), & \mbox{if}~ p_i \geq p_{th}
		\end{cases}.
\end{align}
Note that if $p_{th}\leq 0$, or $p_{th} \geq 1$, $\underline{P_{\sf e}}({\bm p})$ is again  independent of ${\bm p}$, reducing to the previous case. This proves {\bf O5}.

\subsection{Proof of Theorem \ref{thm:eq}}
The only case when $\underline{P_{\sf e}}({\bm p})$ can depend on ${\bm p}$ is when $0 < p_{th} < 1$. In such cases, ${\bm p}$ can simultaneously affect $\underline{P_{\sf e}}({\bm p})$ and $\overline{R}({\bm p})$. 
We next show that it is without loss of generality to consider ${\bm p} = [p_1,\cdots, p_D]$ that has \emph{at most two} distinct values. This will prove that the regions $\mathfrak{R}_1$ and $\mathfrak{R}_2$ are identical and thus prove Theorem \ref{thm:eq}, as $\mathfrak{R}_1$ is obtained from $\mathfrak{R}_2$ by restricting ${\bm p} \in P_{[D]}$  to the smaller set $\mathcal{P}(D)$, i.e., distributions with at most two distinct values.
Given a general ${\bm p} = [p_1,\cdots, p_D]$ that may have more than two distinct values, let $\mathcal{I}_1 = \{i \colon p_i \leq p_{th} \}$, and $\mathcal{I}_2 = [D] \setminus \mathcal{I}_1$.
Consider ${\bm p}' = [p_1',\cdots, p_D']$ such that 
\begin{align} \label{eq:def_pi_prime}
	\sum_{i\in \mathcal{I}_1} p_i = \sum_{i\in \mathcal{I}_1} p_i', ~~ \sum_{i\in \mathcal{I}_2} p_i = \sum_{i\in \mathcal{I}_2} p_i',
\end{align}
and suppose $p_i'$ is uniform over $ \mathcal{I}_1$ (equal to some $p_1' \leq p_{th}$) and is also uniform (equal to some $p_2' > p_{th}$) over $ \mathcal{I}_2$. In other words, $p_i'$ over $\mathcal{I}_1$ is equal to the average value of $p_i$ over $\mathcal{I}_1$, and $p_i'$ over $\mathcal{I}_2$ is equal to the average value of $p_i$ over $\mathcal{I}_2$. Note that $f_1(p)-f_2(p)$ is also monotonic with respect to $p\in [0,1]$. Since $f_1(p_i) - f_2(p_i) \leq 0$ for all $i\in \mathcal{I}_1$ by definition, we have $f_1(p_i') - f_2(p_i') \leq 0$ for all $i\in \mathcal{I}_1$. Similarly, since $f_1(p_i) -f_2(p_i) > 0$ for all $i\in \mathcal{I}_2$, we have $f_1(p_i') - f_2(p_i') > 0$ for all $i\in \mathcal{I}_2$. 

Similar to the concavity argument in the proof of {\bf O2} we have that $\overline{R}({\bm p}) \leq \overline{R}({\bm p}')$. Meanwhile,
\begin{align}
	 \underline{P_{\sf e}}({\bm p}) &= \sum_{i\in [D]} \min \big\{ f_1(p_i), f_2(p_i) \big\} + \min\big\{ \pi_1 \gamma_1, \pi_2 \gamma_2 \big\} \notag \\
	 &= \sum_{i\in \mathcal{I}_1}   f_1(p_i) + \sum_{i\in \mathcal{I}_2}  f_2(p_i)  + \min\big\{ \pi_1 \gamma_1, \pi_2 \gamma_2 \big\} \\
	 &= \sum_{i\in \mathcal{I}_1}   f_1(p_i') + \sum_{i\in \mathcal{I}_2}  f_2(p_i')  + \min\big\{ \pi_1 \gamma_1, \pi_2 \gamma_2 \big\} \label{eq:use_linearity} \\
	 &= \sum_{i\in [D]} \min \big\{ f_1(p_i'), f_2(p_i') \big\} + \min\big\{ \pi_1 \gamma_1, \pi_2 \gamma_2 \big\}  \\
	 &= \underline{P_{\sf e}}({\bm p}'). \label{eq:use_average_and_threshold}
\end{align}
Step \eqref{eq:use_linearity} follows from \eqref{eq:def_pi_prime} and the linearity of $f_1, f_2$. Step \eqref{eq:use_average_and_threshold} is because $f_1(p_i') \leq f_2(p_i')$ if  $i \in \mathcal{I}_1$ and $f_1(p_i') > f_2(p_i')$ if  $i \in \mathcal{I}_2$.
We conclude that ${\bm p}'$ is not restricting the bounds $\overline{R}$ and $\underline{P_{\sf e}}$ and therefore considering ${\bm p}'$ (that has at most two values) will not change the region $\mathfrak{R}_2(\cdot)$. This proves Theorem \ref{thm:eq} \hfill\qed

\section{Proof of Theorem \ref{thm:une_conv}} \label{proof:une_conv}
In this section we only consider \emph{unentangled} protocols. For an unentangled protocol, the initial state $\rho_{A_1B_1A_2B_2\cdots A_TB_T}$ has the form $\sum_{z\in \mathcal{Z}} \lambda_{z} \rho_{A_1\mid z}  \otimes \rho_{B_1\mid z}  \otimes \cdots \otimes \rho_{A_T\mid z}  \otimes \rho_{B_T\mid z} $ for some finite set $\mathcal{Z}$, such that for all $z\in \mathcal{Z}$, $\lambda_{z} \geq 0$,  $\sum_{z\in \mathcal{Z} } \lambda_{z} = 1$, and  $\rho_{A_t \mid z}\in \mathcal{D}(\mathcal{H}_a), \rho_{B_t\mid z}\in \mathcal{D}(\mathcal{H}_b)$ for all $t\in [T]$.

Equivalently, let $Z$ be a random variable over $\mathcal{Z}$ with distribution $\Pr(Z=z) \triangleq P_Z(z) = \lambda_z$ for $z\in \mathcal{Z}$. Consider $ZA_1B_1A_2B_2\cdots A_TB_T$ as a classical-quantum system with density operator
\begin{align}
	&\rho_{ZA_1B_1A_2B_2\cdots A_TB_T} \notag \\
	&= \sum_{z\in \mathcal{Z}}P_Z(z) \ket{z}\bra{z} \otimes   \rho_{A_1\mid z}  \otimes \rho_{B_1\mid z}  \otimes \cdots \otimes \rho_{A_T\mid z}  \otimes \rho_{B_T\mid z}. 
\end{align}
Then the partial state for $A_1B_1\cdots A_TB_T$ matches $\rho_{A_1B_1\cdots A_TB_T}$. 
Note that since the initial state is prepared in advance, $Z$ is independent of the message $W$. The following observations will be useful for the proof.

\begin{enumerate}
	\item Conditioned on $W=w \in [M], Z=z\in \mathcal{Z}$, 
	\begin{align}
		&\omega_{A_1''B_1\cdots A_T''B_T\mid w,z} \notag \\
		&= \omega_{A_1''\mid w,z}\otimes \rho_{B_1\mid z} \otimes \cdots \otimes \omega_{A_T''\mid w,z}\otimes \rho_{B_T\mid z}.\label{eq:unent_tensor_all}
	\end{align}
	\item Thus, conditioned on $Z=z\in \mathcal{Z}$, for $t\in [T]$,
	\begin{align}
		\omega_{WA_t'' B_t   \mid z} = \Big(\frac{1}{M} \sum_{w\in [M]} \ket{w}\bra{w} \otimes \omega_{A_t''\mid w,z}  \Big) \otimes \rho_{B_t\mid z}.\label{eq:unent_conditional_Z_ind_WA_B}
	\end{align}
	\item Conditioned on $Z=z\in \mathcal{Z}$ and $S_t=s \in \{1,2\}$, for $t\in [T]$,
	\begin{align}
		\omega_{A_t''B_t \mid s,z} = \omega_{A_t'' \mid s, z} \otimes \rho_{B_t \mid z}. \label{eq:unent_tensor_AB}
	\end{align}
\end{enumerate}

Let $h_b(P_{\sf c}) = {\sf H}([P_{\sf c},1-P_{\sf c}])$ denote  the binary entropy function. Thus, we have,
{\small
\begin{align}
	& (1-P_{\sf c})\log_2 M-h_b(P_{\sf c})  \notag\\
	& \leq {\sf I}(W;\hat{W}) \label{eq:use_fano} \\
	& \leq {\sf I}(W;A_1''B_1\cdots A_T''B_T)_{\omega} \label{eq:use_holevo} \\
	& \leq {\sf I}(W;A_1''B_1\cdots A_T''B_T \mid Z)_{\omega} \label{eq:cond_Z} \\
	& = {\sf H}(A_1''B_1\cdots A_T''B_T \mid Z)_{\omega} -  {\sf H}(A_1''B_1\cdots A_T''B_T \mid W,Z)_{\omega} \label{eq:use_def_MI} \\
	& \leq \sum_{t \in [T]} {\sf H}(A_t''B_t\mid Z)_{\omega} - {\sf H}(A_1''B_1\cdots A_T''B_T \mid W,Z)_{\omega} \label{eq:use_subadditivity} \\
	& = \sum_{t \in [T]} {\sf H}(A_t''B_t\mid Z)_{\omega} - \sum_{t\in [T]} {\sf H}(A_t''B_t\mid W,Z)_{\omega} \label{eq:use_tensor} \\
	& \leq T\times {\sf I} (W; A_{t^*}''B_{t^*} \mid Z)_{\omega} ~~~~~~ t^* \triangleq \argmax_t {\sf I}(W; A_t''B_t\mid Z)_{\omega} \label{eq:def_t_star} \\
	& = T\times {\sf I}(W; A_{t^*}''  \mid Z)_{\omega} \label{eq:use_condind_Z_WA_B} \\
	& = T\times \sum_{z \in \mathcal{Z} } \lambda_z  \Bigg( {\sf H} \big(\mathcal{N}(\sigma_{A_{t^* }'\mid z})\big) - \frac{1}{M} \sum_{w\in [M]}{\sf H} \big(\mathcal{N}(\sigma_{A_{t^*}'\mid w,z} )\big) \Bigg). \label{eq:bound_message_size}
\end{align}
}Step \eqref{eq:use_fano} uses Fano's inequality. Step \eqref{eq:use_holevo} follows from Holevo's theorem \cite{holevo1973bounds}. Step \eqref{eq:cond_Z} follows from strong subadditivity of von-Neumann entropy, and the independence between $W$ and $Z$. 
Step \eqref{eq:use_def_MI} is by the definition of mutual information. Step \eqref{eq:use_subadditivity} follows from subadditivity of von-Neumann entropy. Step \eqref{eq:use_tensor} is because conditioned on $W=w, Z=z$, $A_1''B_1, A_2''B_2, \cdots A_T''B_T$ are independent as shown by \eqref{eq:unent_tensor_all}.  Step \eqref{eq:use_condind_Z_WA_B} is because $WA_t''$ is independent of $B_t$ given $Z$ (see \eqref{eq:unent_conditional_Z_ind_WA_B}) for all $t\in [T]$ and thus for $t^*$ as well.

Meanwhile  (e.g., from \cite[Ex. 9.1.7]{Wilde_2017}), the detection metric at time slot $t$ is 
{\small
\begin{align}
	P_{\sf d}^t &\geq \frac{1}{2} \Big(1- \Vert \pi_1 \omega_{A_t''B_t \mid S_t=1} - \pi_2 \omega_{A_t''B_t \mid S_t = 2}  \Vert_1 \Big) \\
	&= \frac{1}{2} \Big(1- \Vert \sum_{z \in \mathcal{Z}} \lambda_z \big( \pi_1 \omega_{A_t''B_t \mid S_t=1, z}  - \pi_2 \omega_{A_t''B_t \mid S_t=2, z}  \big) \Vert_1 \Big) \\
	& = \frac{1}{2} \Big(1- \Vert \sum_{z\in \mathcal{Z} } \lambda_z \big( \pi_1 \omega_{A_t''\mid  S_t=1,  z }  - \pi_2 \omega_{A_t'' \mid S_t=2, z}  \big) \otimes \rho_{B_t\mid z}  \Vert_1 \Big)\label{eq:use_condind_Z_At_Bt} \\
	& \geq \frac{1}{2} \Big(1- \sum_{z\in \mathcal{Z} }\lambda_z  \Vert  \big( \pi_1 \omega_{A_t'' \mid S_t=1, z}  - \pi_2 \omega_{A_t'' \mid S_t=2,  z}  \big) \otimes \rho_{B_t\mid z}  \Vert_1 \Big) \label{eq:use_norm_convex} \\
	& = \frac{1}{2} \Big(1- \sum_{z\in \mathcal{Z} }\lambda_z  \Vert  \big( \pi_1 \omega_{A_t''\mid S_t=1,z}   - \pi_2 \omega_{A_t''\mid S_t=2,z}  \big) \Vert_1 \Big) \label{eq:use_norm_tensor}  \\
	&= \sum_{z\in \mathcal{Z} }\lambda_z  \Bigg( \frac{1}{2} \Big(1-  \Vert  \big( \pi_1 \omega_{A_t'' \mid S_t=1, z}   - \pi_2 \omega_{A_t'' \mid S_t=2, z} \big) \Vert_1 \Big) \Bigg) \\
	&= \sum_{z\in \mathcal{Z} }\lambda_z  \Bigg( \frac{1}{2} \Big(1-  \Vert  \big( \pi_1 \mathcal{N}^{(1)}(\sigma_{A_t' \mid z})   - \pi_2 \mathcal{N}^{(2)}(\sigma_{A_t' \mid z}  \big) \Vert_1 \Big) \Bigg).
\end{align}
}Step \eqref{eq:use_condind_Z_At_Bt} is because of \eqref{eq:unent_tensor_AB}.
Step \eqref{eq:use_norm_convex} follows from the convexity of trace norm. Step \eqref{eq:use_norm_tensor} is because $\Vert \rho_{B_t\mid z}  \Vert_1 = 1 ,\forall z \in \mathcal{Z}$ and for normal $A,B$, with eigenvalues $\{\lambda_A^i\}_i$ and $\{\lambda_B^j\}_j$, respectively, we have $\Vert A \Vert_1 \times \Vert B \Vert_1= \Big(\sum_i |\lambda_A^i| \Big) \times \Big(\sum_j |\lambda_B^j|\Big)=  \sum_{i,j} |\lambda_A^i    \lambda_B^j|= \Vert A \otimes B \Vert_1$
where the last step follows from the fact that $\Vert A \otimes B\Vert_1$ is normal and has eigenvalues $\{\lambda_A^i \lambda_B^j\}_{i,j}$. 

Therefore,
{\small
\begin{align} \label{eq:bound_Pd}
	&P_{\sf d} = \max_{t\in [T]} P_{\sf d}^t \notag \\
	& \geq  \sum_{z \in \mathcal{Z}}\lambda_z  \Bigg( \frac{1}{2} \Big(1-  \Vert  \big( \pi_1 \mathcal{N}^{(1)}(\sigma_{A_{t^*}'\mid z})  - \pi_2 \mathcal{N}^{(2)}(\sigma_{A_{t^*}' \mid z} ) \big) \Vert_1 \Big) \Bigg),
\end{align}
}for the same $t^*$ defined in \eqref{eq:def_t_star}.

By definition, if $(R,P_{\sf e})$ is achievable, then for any $\epsilon > 0$, there exists a protocol with $(M,T, P_{\sf c}, P_{\sf d})$ such that $(\log_2 M)/T \geq R-\epsilon, P_{\sf c} < \epsilon, P_{\sf d}\leq P_{\sf e}$. Since \eqref{eq:bound_message_size} and \eqref{eq:bound_Pd} hold for any protocol, it follows that if $(R,P_{\sf e})$ is achievable, then for any $0.5\geq \epsilon > 0$, there exists $M \in \mathbb{N}$, a finite set $\mathcal{Z}$, a set of non-negative real coefficients $\{\lambda_z\}_{z\in \mathcal{Z}}$ such that $\sum_{z\in \mathcal{Z}} \lambda_z = 1$, and a set of density operators $\{\sigma_{A'_{t^*}\mid w,z} \in \mathcal{D}(\mathcal{H}_d) \}_{w\in [M],z\in \mathcal{Z}}$ such that 
\begin{align}
\left\{
\begin{array}{l}
	(1-\epsilon)(R-\epsilon)-h_b(\epsilon)   \leq \mbox{RHS of} ~\eqref{eq:bound_message_size}  \\
	P_{\sf e}  \geq  \mbox{RHS of} ~\eqref{eq:bound_Pd} 
\end{array}
\right..
\end{align}
This implies that any achievable $(R,P_{\sf e})$ is inside
{\small
\begin{align}
	 \conv \left\{\begin{array}{l} (R,P_{\sf e}) \colon \\ \exists M \in \mathbb{N}, ~~\sigma_1,\sigma_2,\cdots,\sigma_M \in \mathcal{D}(\mathcal{H}_d),  \\~ \sigma \triangleq \frac{1}{M}\sum_{w\in [M]}  \sigma_w, \\ 
	R \leq {\sf H}(\mathcal{N}(\sigma)) - \frac{1}{M}\sum_{w\in [M]} {\sf H}(\mathcal{N}(\sigma_w)), \\ P_{\sf e} \geq \frac{1}{2}\Big( 1-\Vert \pi_1 \mathcal{N}^{(1)}(\sigma) - \pi_2 \mathcal{N}^{(2)}(\sigma) \Vert_1 \Big) \end{array} \right\}.
\end{align}
}
This concludes the proof of Theorem \ref{thm:une_conv}.
\hfill\qed

\section{Proof of Theorem \ref{thm:unentangled}} \label{proof:unentangled}
\subsection{Converse proof for Theorem \ref{thm:unentangled}}
We first prove $\mathfrak{R}_u^* \subseteq \mathfrak{R}(d)$ for the IDE channels defined in Section \ref{sec:special_channel}. The proof starts by applying Theorem \ref{thm:une_conv}.  Let $\sigma = UP U^\dagger$, $P = \diag([p_1,p_2,\cdots, p_{d}])$ be the spectral decomposition of $\sigma$. Therefore $0\leq p_i \leq 1, \forall i\in [d]$ and $p_1 +\cdots + p_{d}  = 1$. We have,
\begin{align}
	 \mathcal{N}(\sigma)  &=  \underbrace{(\theta_1\alpha_1 +\theta_2\alpha_2)}_{\bar{\alpha}}\sigma  ~\boxplus~  \underbrace{(\theta_1\beta_1+\theta_2\beta_2)}_{\bar{\beta}} I_d/d \notag \\
	 &~~~~~~~~~~~~~~~~~~~~~~~~~\boxplus~   \underbrace{(\theta_1\gamma_1 + \theta_2\gamma_2)}_{\bar{\gamma}} \ket{0}\bra{0} \\
	 &= \bar{\alpha} U P U^\dagger ~\boxplus~   \bar{\beta} UU^\dagger/d ~\boxplus~   \bar{\gamma} \ket{0}\bra{0},
\end{align}
and thus,
\begin{align}
	  &{\sf H}(\mathcal{N}(\sigma)) \notag \\
	  &=  H([p_1 \bar{\alpha}+ \bar{\beta}/d, ~  p_2 \bar{\alpha}+ \bar{\beta}/d,   ~\cdots, p_{d} \bar{\alpha}+ \bar{\beta}/d, ~\bar{\gamma}]).
\end{align}
Similarly, for $w\in [M]$, let $\sigma_w = U_w\diag([p_1^{(w)},\cdots, p_{d}^{(w)}])U_w^\dagger$. Note that $0\leq p_i^{(w)}\leq 1, \forall i\in [d]$ and that  $p_1^{(w)}+\cdots + p_{d}^{(w)} = 1$. We have,
{\small
\begin{align}
	&{\sf H}(\mathcal{N}(\sigma_w)) \notag \\
	&= {\sf H}([p_1^{(w)}\bar{\alpha}+ \bar{\beta}/d,   ~~p_2^{(w)}\bar{\alpha}+ \bar{\beta}/d,  ~~ \cdots, p_{d}^{(w)}\bar{\alpha}+ \bar{\beta}/d, ~~\bar{\gamma}]) \\
	& \geq p_1^{(w)}{\sf H}([\bar{\alpha}+ \bar{\beta}/d, \bar{\beta}/d, \cdots, \bar{\beta}/d,\bar{\gamma}]) \notag \\
	&~~+ p_2^{(w)}{\sf H}([\bar{\beta}/d, \bar{\alpha}+ \bar{\beta}/d, \cdots, \bar{\beta}/d,\bar{\gamma}])\notag\\
	&~~+\cdots + p_{d}^{(w)}{\sf H}([\bar{\beta}/d, \bar{\beta}/d, \cdots, \bar{\alpha}+ \bar{\beta}/d,\bar{\gamma}]) \label{eq:conv_use_convex} \\
	& = (p_1^{(w)} + \cdots + p_{d}^{(w)}){\sf H}([\bar{\alpha}+ \bar{\beta}/d, \bar{\beta}/d, \cdots, \bar{\beta}/d,\bar{\gamma}]) \\
	& = {\sf H}([\bar{\alpha}+ \bar{\beta}/d, \bar{\beta}/d, \cdots, \bar{\beta}/d,\bar{\gamma}]).
\end{align}
}Step \eqref{eq:conv_use_convex} follows from the concavity of entropy. Therefore,
{\small
\begin{align} 
	R &\leq {\sf H}([p_1\bar{\alpha}+ \bar{\beta}/d,   p_2 \bar{\alpha}+ \bar{\beta}/d,   \cdots, p_{d} \bar{\alpha}+ \bar{\beta}/d, \bar{\gamma}]) \notag \\
	&~~~~-{\sf H}([\bar{\alpha}+ \bar{\beta}/d, \bar{\beta}/d, \cdots, \bar{\beta}/d,\bar{\gamma}])  \\
	& = -\sum_{i\in [d]}\big(\bar{\alpha} p_i +\bar{\beta}/d \big)  \log_2 \big(\bar{\alpha}p_i +\bar{\beta}/d \big) \notag \\
	&~~+ (\bar{\alpha}+\bar{\beta}/d)\log_2 (\bar{\alpha}+\bar{\beta}/d) + (d-1) (\bar{\beta}/d) \log_2 (\bar{\beta}/d). \label{eq:rate_bound_1}
\end{align}
}Meanwhile,
{\small
\begin{align}
	& P_{\sf e} \geq \frac{1}{2}\Big( 1- \Vert \pi_1 \mathcal{N}^{(1)}(\sigma) - \pi_2 \mathcal{N}^{(2)}(\sigma) \Vert_1 \Big) \\
	& = \frac{1}{2}\Big( 1- \Vert (\pi_1\alpha_1- \pi_2\alpha_2)\sigma + (\pi_1\beta_1-\pi_2\beta_2 ) I_d/d  \notag \\
	&~~~~~~~~~~~~~~~~~~~~~~~~~~~~~~~~~~~+ (\pi_1\gamma_1 -\pi_2\gamma_2)\ket{0}\bra{0}  \Vert_1 \Big) \\
	& = \frac{1}{2}\Big( 1- \Vert  U (\pi_1\alpha_1- \pi_2\alpha_2)P U^\dagger + U(\pi_1\beta_1-\pi_2\beta_2)/d U^\dagger \notag \\
	&~~~~~~~~~~~~~~~~~~~~~~~~~~~~~~~~~~~+ (\pi_1\gamma_1 -\pi_2\gamma_2)\ket{0}\bra{0}  \Vert_1 \Big) \\
	& = \frac{1}{2}\Big( 1 - \sum_{i\in [d]}\big( \vert (\pi_1 \alpha_1 -\pi_2\alpha_2)p_i + (\pi_1\beta_1-\pi_2\beta_2)/d  \vert\big) \notag \\
	&~~~~~~~~~~~~~~~~~~~~~~~~~~~~~~~~~~~~~~~~~~~~~~ - \vert \pi_1\gamma_1 -\pi_2\gamma_2 \vert  \Big) \label{eq:use_SVD} \\
	&= \frac{1}{2}\Big(1+2\sum_{i\in [d]}\min  \big\{ \pi_1 \alpha_1 p_i + \pi_1 \beta_1/d  , ~  \pi_2 \alpha_2 p_i + \pi_2 \beta_2/d  \big\} \notag \\
	&~~~~~~~~~~~~~~~~ - \sum_{i\in [d]}(\pi_1 \alpha_1 p_i + \pi_1 \beta_1/d +  \pi_2 \alpha_2 p_i + \pi_2 \beta_2/d )  \notag\\
	&~~~~~~~~~~~~~~~~ +2\min\big\{ \pi_1 \gamma_1, \pi_2 \gamma_2 \big\}-(\pi_1\gamma_1+\pi_2\gamma_2)\Big)  \label{eq:De_bound_1}\\
	& = \sum_{i\in [d]}\min  \big\{ \pi_1 \alpha_1 p_i + \pi_1 \beta_1/d  , ~  \pi_2 \alpha_2 p_i + \pi_2 \beta_2/d  \big\} \notag \\
	&~~~~~~~~~~~~~ + \min\big\{ \pi_1 \gamma_1, \pi_2 \gamma_2 \big\}. \label{eq:De_bound_2}	
\end{align}
}Step \eqref{eq:use_SVD} is because $\Vert \bar{U} A\bar{U}^\dagger \Vert_1   = \Vert A \Vert_1$ for any diagonal matrix $A$ and unitary matrix $\bar{U}$ and we consider $\bar{U} = \bbsmatrix{U & 0\\0 & 1}$.
To see Step \eqref{eq:De_bound_1}, note that for any two real numbers $u,v$,  we have $|u -v|  = u +v- 2\min\{u, v\}$. Note that $p_i$ can be considered as the value of $P_X(i)$ for $i\in [d]$, where $P_X \in P_{[d]}$. According to Theorem \ref{thm:une_conv}, we obtain that,
{\small
\begin{align} \label{eq:region_1}
	\mathfrak{R}^*_u  &\subseteq 
	\conv \left\{\begin{array}{l}  (R,P_{\sf e}) \colon  \\ \exists P_X   \in P_{[d]}, p_i = P_X(i), \forall i\in [d] \\ R \leq  \mbox{RHS of}~ \eqref{eq:rate_bound_1}  \\ P_{\sf e} \geq \mbox{RHS of}~ \eqref{eq:De_bound_2} \end{array} \right\} \\
	 &= \mathfrak{R}_2(d, \alpha_{[2]}, \beta_{[2]}, \gamma_{[2]}, \theta_{[2]},  \pi_{[2]}).
\end{align}}
\hfill\qed

\subsection{Achievability proof for Theorem \ref{thm:unentangled}}
We next show that $\mathfrak{R}^*_u \supseteq  \mathfrak{R}_2(d, \alpha_{[2]}, \beta_{[2]}, \gamma_{[2]}, \theta_{[2]}, \pi_{[2]})$. From Remark \ref{rem:com_rdm} we know that the framework allows arbitrary common randomness shared between the transmitter and the receiver.  Let $Z$ denote this random variable shared between the transmitter and the receiver. The distribution of $Z$ is to be designed later. In the following we will use a purely classical strategy for the proof of achievability, and therefore the protocol considered is unentangled.
Given a distribution $P_X \in P_{[d]}$, and $M,T \in \mathbb{N}$, let $\mathfrak{C}^{M\times T}$ be the set of $M\times T$ matrices with elements in $[d]$. Each element of $\mathfrak{C}^{M\times T}$  is referred to as a codebook. The transmitter and the receiver  use the common randomness $Z$ to pick  a random codebook $\mathcal{C}$. To do this, let $X_{w,t}$ be the $(w,t)^{th}$ entry of $\mathcal{C}$. Then $X_{w,t}$ is generated independently for $w\in [M], t\in [T]$, with distribution determined as $\Pr(X_{w,t} = x) = P_X(x)$ for $x\in [d]$ and for each $X_{w,t}$. Therefore, $Z$ should have $\mathcal{C}$ as one of its components.
Suppose the transmitter and the receiver pick the codebook $\mathcal{C}=C$.  Let $x_{w,t}$ denote the $(w,t)^{th}$ entry of $C$. The encoding process is such that for message $w \in [M]$, the transmitter at time slot $t\in [T]$ sets $\bar{A}_t'$ to the state $\sigma_{\bar{A}_t'\mid x_{w,t}} = \ket{x_{w,t}}\bra{x_{w,t}}$. $\bar{A}_t'$ then goes through the channel $\mathcal{N}^{(S_t)}$, so that $A_t'$ evolves into $A_t''$. Further conditioned on $S_t=s$ for $s\in \{1,2\}$,  $A_t''$ is in the mixed state,
\begin{align}
	&\omega_{A_t'' \mid x_{w,t}, S_t=s} \notag \\
	&= \alpha_s \ket{x_{w,t}}\bra{x_{w,t}} ~\boxplus~  (\beta_s /d)I_{d}   ~\boxplus~  \gamma_s \ket{0}\bra{0}  \\
	&= \alpha_s \ket{x_{w,t}}\bra{x_{w,t}} ~\boxplus~   ( \beta_s/d) \sum_{k\in [d]} \ket{k}\bra{k}   ~\boxplus~  \gamma_s \ket{0}\bra{0} \\
	& = (\alpha_s + \beta_s/d) \ket{ x_{w,t} }\bra{ x_{w,t} } \notag \\
	&~~~~~~\boxplus~  \sum_{k \in [d], k \not=  x_{w,t}} (\beta_s/d)  \ket{k}\bra{k}  ~\boxplus~  \gamma_s \ket{0}\bra{0}. \label{eq:output_state_une}
\end{align}
The receiver at time slot $t$  measures $A_t''\bar{B}_t$ on the PVM $\{\ket{y}\bra{y}\}_{y \in [0:d]}$, 
with the classical output labeled as $Y_t = y \in [0:d]$. ($Y_t=0$ means the received system is erased.)
According to \eqref{eq:output_state_une}, the conditional probability 
\begin{align} \label{eq:classical_cond_dist_une}
	& \Pr(Y_t= y \mid  X_{w,t} = x, S_t = s) \notag \\
	& = \begin{cases}
		\alpha_s + \beta_s/d, & y = x \\
		\beta_s/d, & y = x', ~\forall x' \in [d] \setminus \{x\} \\
		\gamma_s, & y=0,
	\end{cases},
\end{align}
for $x\in [d],s\in \{1,2\}, y\in [0:d]$.  Since $X_{w,t}$ is independent of $S_t$, it follows that
\begin{align}
	& \Pr(Y_t = y\mid S_t=s) \notag \\
	& =  \sum_{x\in [d]}\Pr(X_{w,t} = x) \Pr(Y_t= y \mid  X_{w,t} = x, S_t = s) \\
	& =  \sum_{x\in [d]}P_X(x) \Pr(Y_t= y \mid  X_{w,t} = x, S_t = s) \\
	& = \begin{cases}
		\alpha_s P_X(y) + \beta_s/d, & y \in [d] \\
		\gamma_s, & y=0
	\end{cases},
\end{align}
for $y\in [0:d]$.
 
Given $Y_t=y$, let the detection rule be
{\small
\begin{align} \label{eq:MAP_detector_une}
	&\hat{S}_t(y) \notag \\
	&= \begin{cases}
		1, & \mbox{if} ~\pi_1 \Pr(Y_t = y  \mid S_t = 1) \geq \pi_2 \Pr(Y_t = y \mid S_t = 2) \\
		2, & \mbox{if} ~ \pi_1 \Pr(Y_t = y  \mid S_t = 1) < \pi_2 \Pr(Y_t = y \mid S_t = 2)
	\end{cases},
\end{align}
}which is the MAP detector with respect to prior distribution $(\pi_1, \pi_2)$. We obtain the detection metric at time slot $t$,
{\small
\begin{align}
	&P_{\sf d}^t  = \sum_{y\in  [0:d] } \min_{s\in \{1,2\}} \big\{\pi_s \Pr(Y_t = y  \mid S_t = s)   \big\} \\
	& = \sum_{y\in [d]}\min  \big\{  \pi_1 \alpha_1 P_X(y) + \pi_1 \beta_1/d   , ~  \pi_2 \alpha_2 P_X(y) + \pi_2 \beta_2/d  \big\} \notag \\
	&~~~~~~~~~~~~~+ \min\big\{ \pi_1 \gamma_1, \pi_2 \gamma_2 \big\}.  \label{eq:achievable_detection_error_une}
\end{align}
}Since $P_{\sf d}^t$ is the same across $t$, the worst case over $T$  time slots is equal to the average case, $P_{\sf d} = P_{\sf d}^t$. 

Next, for the communication task, again since $X_{w,t}$ is independent of $S_t$, it follows from \eqref{eq:classical_cond_dist_une} that,
{\small
\begin{align}
	& \Pr(Y_t = y \mid X_{w,t} = x) \notag \\
	& = \sum_{s\in \{1,2\}} \Pr(S_t = s) \Pr(Y_t= y \mid X_{w,t} =x, S_t=s) \\
	& = \begin{cases}
		\theta_1( \alpha_1 + \beta_1/d ) + \theta_2( \alpha_2 + \beta_2/d ), & y = x \\
		\theta_1\beta_1/d + \theta_2\beta_2/d, & y = x', ~\forall x' \in [d] \setminus \{x\} \\
		\theta_1\gamma_1 + \theta_2\gamma_2 & y=0
	\end{cases} \\
	& = \begin{cases}
		\bar{\alpha} + \bar{\beta}/d, & y = x \\
		\bar{\beta}/d, & y = x', ~\forall x' \in [d] \setminus \{x\} \\
		\bar{\gamma} & y=0
	\end{cases}. \label{eq:classical_cond_dist_YX_une}
\end{align}
}Consider a classical channel $P_{Y\mid X}(y\mid x)$ which has the same conditional probability distribution as $\Pr(Y_t = y \mid X_{w,t} = x)$. 
From classical information theory, if $(\log_2 M)/T < {\sf I}(X;Y)_{P_{Y\mid X}P_X} = -\sum_{i\in [d]}\big(\bar{\alpha} P_X(i) +\bar{\beta}/d \big)  \log_2 \big(\bar{\alpha}P_X(i) +\bar{\beta}/d \big) + (\bar{\alpha}+\bar{\beta}/d)\log_2 (\bar{\alpha}+\bar{\beta}/d) + (d-1) (\bar{\beta}/d) \log_2 (\bar{\beta}/d)$, then there exists a  decoder with $P_{\sf c}$ arbitrarily small as $T \to \infty$. Together with \eqref{eq:achievable_detection_error_une},  this shows that any $(R,P_{\sf e})$ in 
{\small
\begin{align} \label{eq:region_wo_conv_une}
	\left\{\begin{array}{l} (R,P_{\sf e}) \colon \\\exists P_X \in P_{[d]} \\
		R \leq -\sum_{i\in [d]}\big(\bar{\alpha} P_X(i) +\bar{\beta}/d \big) \log_2 \big(\bar{\alpha}P_X(i)+\bar{\beta}/d \big) \notag \\ ~~~~~~+ (\bar{\alpha}+\bar{\beta}/d)\log_2 (\bar{\alpha}+\bar{\beta}/d) + (d-1) (\bar{\beta}/d)\log_2 (\bar{\beta}/d)   \\
		P_{\sf e} \geq \sum_{i\in [d]}\min_{s\in \{1,2\}}  \big\{  \pi_s \alpha_s P_X(i) + \pi_s \beta_s/d  \big\} \notag \\ ~~~~~~ +\min\big\{ \pi_1 \gamma_1, \pi_2 \gamma_2 \big\}
		\end{array} \right\}
\end{align}
}%
is achievable. 

Finally, given two series of such protocols that achieve $(R^{(1)}, P_{\sf e}^{(1)})$ and $(R^{(2)}, P_{\sf e}^{(2)})$, respectively, by concatenating them in time and interleaving the time slots (in a random manner), it can be shown that for any positive $\lambda_1, \lambda_2$, $\lambda_1+\lambda_2 = 1$,
\begin{align}
	(\lambda_1 R^{(1)}+ \lambda_2 R^{(2)}, \lambda_1P_{\sf e}^{(1)}+ \lambda_2P_{\sf e}^{(2)}),
\end{align}
is achievable as well. This accounts for the convex hull operation. Therefore, we conclude that any
\begin{align} \label{eq:region_une_last_equation}
	(R, P_{\sf e}) \in \mathfrak{R}_2(d, \alpha_{[2]}, \beta_{[2]}, \gamma_{[2]},\theta_{[2]},\pi_{[2]}),
\end{align}
is achievable by unentangled protocols.  This concludes the proof of Theorem \ref{thm:unentangled}. 
\hfill\qed

\section{Proofs of Theorem \ref{thm:superdense_coding} and Theorem \ref{thm:unreliable}} \label{proof:superdense_coding}
To prove these theorems, we first  recall the superdense coding protocol \cite{Superdense}, \cite[Sec. 6.5.2]{Wilde_2017}.

\noindent{\it Superdense coding:} 
Given $d \in \mathbb{N}$, for $(i,j) \in [0:d-1]^2$, define $\ket{\phi_{i,j}} \in \mathcal{H}_d^A\otimes \mathcal{H}_d^B$ as the state for two $d$-ary  quantum systems $A,B$, such that,
\begin{align} \label{eq:bell_basis}
\begin{cases}
	\ket{\phi_{0,0}} = \frac{1}{\sqrt{d}}  \big(\ket{1}_A\ket{1}_B + \ket{2}_A\ket{2}_B + \cdots + \ket{d}_A\ket{d}_B\big)  \\
	\ket{\phi_{i,j}} = ({\sf X}^{i} {\sf Z}^{j} \otimes I_d) \ket{\phi_{0,0}}, \forall (i,j) \in [0:d-1]^2 
\end{cases},
\end{align}
where $\{\ket{1}_A,\cdots,\ket{d}_A\}$, $\{\ket{1}_B,\cdots, \ket{d}_B\}$ denote the sets of computational basis vectors for $\mathcal{H}_d^A$ and $\mathcal{H}_d^B$, respectively. ${\sf X}$ and ${\sf Z}$ are the d-ary Pauli X and Pauli Z gates, defined such that $\forall k\in [d]$,
\begin{align}
	{\sf X}\ket{k} = 
	\begin{cases}
		\ket{k+1}, & \forall k \in [d-1]\\
		\ket{1}, & k = d
	\end{cases},
	&&{\sf Z}\ket{k} = e^{i2\pi (k-1)/d} \ket{k}.
\end{align}
It is known (e.g., see \cite[Sec. 6.5.2]{Wilde_2017}) that $\{\ket{\phi_{i,j}}\}_{(i,j)\in [0:d-1]^2}$ constitute a set of orthonormal basis vectors for $\mathcal{H}^A_d\otimes \mathcal{H}^B_d$ and each of the states  $\ket{\phi_{i,j}}$ is maximally entangled.

\subsection{Proof of Theorem \ref{thm:superdense_coding}}
 From Remark \ref{rem:com_rdm} we know that the framework allows arbitrary common randomness shared between the transmitter and the receiver. We choose the initial state shown in Remark \ref{rem:com_rdm} as,
 {\small
\begin{align}
	&\rho_{A_1B_1\cdots A_TB_T} = \rho_{Z_{A_1}\bar{A}_1 \bar{B}_1Z_{B_1} \cdots Z_{A_T}\bar{A}_T \bar{B}_TZ_{B_T}}  \\
	& =  \sum_{z\in \mathcal{Z}}P_{Z}(z) \bigotimes_{t=1}^T\Big( \ket{z}\bra{z}_{Z_{A_t}}  \otimes\ket{\phi_{0,0}}\bra{\phi_{0,0}}_{\bar{A}_t\bar{B}_t} \otimes\ket{z}\bra{z}_{Z_{B_t}}\Big).
\end{align}
}In other words, such a protocol starts by distributing $T$ copies of the pure state $\ket{\phi_{0,0}}$ to the transmitter and the receiver, and allows the two parties to share the classical common randomness $Z$ (to be designed later).

With this we are ready to construct the protocol. 
For $x\in [d^2]$, let
\begin{align}
	i_x \triangleq \lfloor (x-1)/d \rfloor, ~j_x \triangleq (x-1) \mod d,
\end{align}
so that there is a one-to-one correspondence between $x \leftrightarrow  (i_x, j_x)$, where $x\in [d^2]$ and $(i_x,j_x) \in [0:d-1]^2$.  Given a distribution $P_X \in P_{[d^2]}$, and $M,T \in \mathbb{N}$, let $\mathfrak{C}^{M\times T}$ be the set of $M\times T$ matrices with elements in $[d^2]$.  Proceeding similarly as in the achievability proof of Theorem \ref{thm:unentangled}, each element of $\mathfrak{C}^{M\times T}$  is referred to as a codebook, and  a random codebook  $\mathcal{C}$ is chosen by the transmitter and receiver based on their common randomness $Z$.  Suppose the instance $\mathcal{C}=C$ is chosen.   Let $x_{w,t}$ denote the $(w,t)^{th}$ entry of $C$. For message $w \in [M]$, the transmitter at time slot $t\in [T]$  applies the unitary operator ${\sf X}^{i_{x_{w,t}}}{\sf Z}^{j_{x_{w,t}}}$ to $\bar{A}_t$, so that $\bar{A}_t\bar{B}_t$ evolves to $\bar{A}_t'\bar{B}_t$  and the state is $\sigma_{\bar{A}_t'\bar{B}_t\mid x_{w,t}} = \ket{\phi_{i_{x_{w,t}},j_{x_{w,t}}}}\bra{\phi_{i_{x_{w,t}},j_{x_{w,t}}}}$ according to \eqref{eq:bell_basis}. Let the output system of $\mathcal{E}_t$ be $A_t' = \bar{A}_t'$, which then goes through the channel $\mathcal{N}^{(S_t)}$, so that $A_t'\bar{B}_t$ evolves into $A_t''\bar{B}_t$. Further conditioned on $S_t=s$ for $s\in \{1,2\}$,  $A_t''\bar{B}_t$ is in the mixed state,
\begin{align}
	&\omega_{A_t''\bar{B}_t\mid x_{w,t}, S_t=s} \notag \\
	&= \alpha_s \ket{\phi_{i_{x_{w,t}},j_{x_{w,t}}}}\bra{\phi_{i_{x_{w,t}},j_{x_{w,t}}}} \boxplus (\beta_s/d^2) I_{d^2} \boxplus \gamma_s \ket{0}\bra{0} \label{eq:output_state_inter} \\ 
	&= \alpha_s \ket{\phi_{i_{x_{w,t}},j_{x_{w,t}}}}\bra{\phi_{i_{x_{w,t}},j_{x_{w,t}}}} \notag \\
	&~~~~~~\boxplus  ( \beta_s/d^2) \sum_{(k,l)\in [0:d-1]^2} \ket{\phi_{k,l}}\bra{\phi_{k,l}}  \boxplus \gamma_s \ket{0}\bra{0} \\
	& = (\alpha_s + \beta_s/d^2) \ket{\phi_{i_{x_{w,t}},j_{x_{w,t}}}}\bra{\phi_{i_{x_{w,t}},j_{x_{w,t}}}} \notag \\
	&~\boxplus \sum_{(k,l)\in [0:d-1]^2, (k,l) \not= (i_{x_{w,t}},j_{x_{w,t}})} \hspace{-1cm} (\beta_s/d^2)  \ket{\phi_{k,l}}\bra{\phi_{k,l}} \boxplus \gamma_s \ket{0}\bra{0}. \label{eq:output_state}
\end{align}
The receiver at time slot $t$  measures $A_t''\bar{B}_t$ on the PVM $\{\ket{\phi_{k,l}}\bra{\phi_{k,l}}\}_{(k,l) \in [0:d-1]^2} \cup \{\ket{0}\bra{0}\}$, 
with the classical output labeled as $Y_t = y \in [0:d^2]$  such that $(i_y, j_y) = (k,l)$ if $A_t''$ is not erased, and $Y_t=0$ if $A_t''$ is erased.
According to \eqref{eq:output_state}, the conditional probability,
\begin{align} \label{eq:classical_cond_dist}
	 &\Pr\Big(Y_t= y \mid  X_{w,t} = x, S_t = s\Big)\notag \\
	  & = \begin{cases}
		\alpha_s + \beta_s/d^2, & y = x \\
		\beta_s/d^2, & y = x', ~\forall x' \in [d^2] \setminus \{x\} \\
		\gamma_s, & y=0,
	\end{cases},
\end{align}
for $x\in [d^2],s\in \{1,2\}, y\in [0:d^2]$.  
We observe that the conditional distribution obtained in \eqref{eq:classical_cond_dist} is equal to that in  \eqref{eq:classical_cond_dist_une} by substituting $d$ with $d^2$. Similar to the arguments from \eqref{eq:classical_cond_dist_une} to \eqref{eq:region_une_last_equation} (construction of the MAP detector \eqref{eq:MAP_detector_une}, existence of the decoder, time-sharing), we can show that any $(R,P_{\sf e})$ in $\mathfrak{R}_2(d^2, \alpha_{[2]}, \beta_{[2]}, \gamma_{[2]}, \theta_{[2]}, \pi_{[2]})$ is achievable. \hfill \qed

\subsection{Proof of Theorem \ref{thm:unreliable}} 
It suffices to show the achievability of $\mathfrak{R}_{sdc}$ in the setting of unreliable entanglement because the rest of the proof follows from the achievability of $\mathfrak{R}_u^*$ and time-sharing.
Based on the superdense coding protocol in the proof of Theorem \ref{thm:superdense_coding} (Section \ref{proof:superdense_coding}), now consider that the quantum systems $\bar{B}_1,\bar{B}_2,\cdots, \bar{B}_T$ undergo $\mathcal{N}_B$. It will be convenient to assume that $\mathcal{N}_B$ are applied before the  superdense coding  protocol is initiated. 

Note that 
\begin{align}
	\mathcal{N}_B(\rho) = \widetilde{\alpha} \rho + \widetilde{\beta} I_{d}/d. 
\end{align}
Then instead of $\rho_{\bar{A}_t\bar{B}_t} = \ket{\phi_{0,0}}\bra{\phi_{0,0}}$  as shown earlier in the proof of Theorem \ref{thm:superdense_coding}, we now have
{\small 
\begin{align}
	\rho_{\bar{A}_t\bar{B}_t} = (I_d \otimes \mathcal{N}_B)(\ket{\phi_{0,0}}\bra{\phi_{0,0}}) = \widetilde{\alpha}\ket{\phi_{0,0}}\bra{\phi_{0,0}} + \widetilde{\beta}I_{d^2}/d^2,
\end{align}
}and thus conditioned on $x_{w,t}$,
\begin{align}
	\sigma_{\bar{A}_t'\bar{B}_t\mid x_{w,t}} =   \widetilde{\alpha}\ket{\phi_{i_{x_{w,t}},j_{x_{w,t}}}}\bra{\phi_{i_{x_{w,t}},j_{x_{w,t}}}} + \widetilde{\beta}I_{d^2}/d^2.
\end{align}
Similar to \eqref{eq:output_state_inter} we have that
{\small
\begin{align}
	&\omega_{A_t''\bar{B}_t\mid x_{w,t}, S_t=s} \notag \\
	& =  \alpha_s \sigma_{\bar{A}_t'\bar{B}_t\mid x_{w,t}} \boxplus (\beta_s/d^2) I_{d^2} \boxplus \gamma_s \ket{0}\bra{0} \\
	& = \underbrace{\alpha_s \widetilde{\alpha}}_{\ddot{\alpha}_s} \ket{\phi_{i_{x_{w,t}},j_{x_{w,t}}}}\bra{\phi_{i_{x_{w,t}},j_{x_{w,t}}}} \boxplus \underbrace{(\alpha_s \widetilde{\beta} + \beta_s)}_{\ddot{\beta}_s}/d^2 \boxplus \gamma_s\ket{0}\bra{0},
\end{align}
}which is \eqref{eq:output_state_inter} with $\alpha_s, \beta_s$ replaced as $\ddot{\alpha}_s, \ddot{\beta}_s$. The rest of the proof follows from that of Theorem \ref{thm:superdense_coding}. It also follows from the result of Theorem \ref{thm:superdense_coding} that any $(R,P_{\sf e})$ in $\mathfrak{R}_{sdc}$ is achievable with unreliable entanglement. \hfill \qed

\bibliographystyle{IEEEtran}
\bibliography{../../bib_file/yy.bib}
\end{document}